\documentclass[prx,aps,superscriptaddress,twocolumn,amsmath,amssymb,floatfix]{revtex4-2} 

\usepackage[breaklinks,colorlinks = true,linkcolor = magenta,urlcolor=magenta,citecolor=red,hypertexnames=false]{hyperref}
\usepackage{graphicx}
\usepackage{multirow}
\usepackage{graphicx}% Include figure files
\usepackage{dcolumn}% Align table columns on decimal point
\usepackage{bm}% bold math
\usepackage{color}
\usepackage{subfigure}

\usepackage{tabularx}
\usepackage[dvipsnames]{xcolor}
\usepackage{lipsum}
\usepackage{braket}
\usepackage{float}
\usepackage{silence}
\allowdisplaybreaks
\def\be{\begin{equation}}
\def\ee{\end{equation}}

\makeatletter
\newsavebox{\@brx}
\newcommand{\llangle}[1][]{\savebox{\@brx}{\(\m@th{#1\langle}\)}%
  \mathopen{\copy\@brx\mkern2mu\kern-0.9\wd\@brx\usebox{\@brx}}}
\newcommand{\rrangle}[1][]{\savebox{\@brx}{\(\m@th{#1\rangle}\)}%
  \mathclose{\copy\@brx\mkern2mu\kern-0.9\wd\@brx\usebox{\@brx}}}
\makeatother

\begin{document}
      \title{Stress Response of Jammed Solids: Prestress and Screening}
\author{Surajit Chakraborty}
\affiliation{Tata Institute of Fundamental Research, Hyderabad 500046, India}
\author{Jishnu N. Nampoothiri}
\affiliation{Tata Institute of Fundamental Research, Hyderabad 500046, India}
\affiliation{Martin Fisher School of Physics, Brandeis University, Waltham, Massachusetts 02454, USA}
\author{Subhro Bhattacharjee}
\affiliation{International Centre for Theoretical Sciences, Tata Institute of Fundamental Research, Bengaluru 560089, India}
\author{Bulbul Chakraborty}
\affiliation{Martin Fisher School of Physics, Brandeis University, Waltham, Massachusetts 02454, USA}
\author{Kabir Ramola }
\affiliation{Tata Institute of Fundamental Research, Hyderabad 500046, India}

\date{\today}

%%%%%%%
\begin{abstract}
Unlike classical elasticity, where stresses arise from deformations relative to a stress-free reference configuration, rigidity in amorphous systems is maintained by disordered force networks that generate internal prestress. Previously, we introduced a ``stress-only'' formulation, where mechanical equilibrium resembles Gauss's law in a rank-2 tensor electrostatics with vector charges, and demonstrated that the mechanical response of jammed solids is described by the dielectric response of this gauge-theoretic formulation. Here, we extend this framework by incorporating scale-dependent screening that captures both dielectric and Debye-type behaviour. This introduces a characteristic length scale in stress correlations as well as in the response to external forces. Through numerical simulations of soft-sphere packings, we show that this length scale is set by the particle size, thus providing a natural ultraviolet cutoff while preserving long-wavelength emergent elasticity. We show that this lengthscale remains finite for all pressures, with no evidence for an emergent Debye-like screening near the frictionless unjamming transition. We demonstrate that although individual realisations show strong fluctuations, disorder averaging at fixed macroscopic conditions yields a robust dielectric-like response that persists up to unjamming. Finally, we also provide a physical interpretation of the gauge field within the electrostatic mapping: relative grain displacements in response to localised external perturbations correspond to difference in the gauge field, linking the field-theoretic description to particle-level mechanics.
\end{abstract}
\maketitle

\section{Introduction}
The theory of classical elasticity of solids describes their mechanical responses to deforming forces. It assumes that internal forces, manifested as the stress tensor, arise from displacements relative to a well-defined, stress-free reference state. These forces are proportional to the strain field -- calculated from the derivative of the displacement from the reference state, with the proportionality constant - the elastic modulus - characterising the rigidity of the solid in the linear regime~\cite{landau2012theory}. Canonical examples are crystalline solids where the spontaneous symmetry breaking provides a unique zero stress reference state with respect to which the strain field is meaningfully defined, along with the right number of elastic moduli, with the structure of the latter tied to the nature of the symmetry breaking via the space-group~\cite{landau2012theory}.

Many soft solids~\cite{alexander1998amorphous,cates1998jamming,bouchaud2004course,bi2011jamming,brown2014shear,lemaitre2017inherent,tong2020emergent,zhang2022prestressed} exhibit structural and mechanical features that emerge from ``frozen-in'' stresses, or prestress that arise from a variety of sources. Jammed solids gain their rigidity from external forces, and the resulting prestress is essential in defining their solidity~\cite{cates1998jamming,pica2009jamming,bi2011jamming}. Colloidal and polymer gels form in processes where thermal fluctuations of the constituents are frozen, and the internal stresses that are generated can at best be partially redistributed~\cite{Mao2009,bouzid_elastically_2017,delgado2021}. Common to all these cases is the absence of a unique, stress-free reference state: prestress determines the mechanical response, which, in turn, influences the assembly of the stress-bearing network.  This nonlinear feedback cannot be rigorously incorporated within the framework of classical elasticity, where strain, defined with respect to a stress-free reference structure, leads to stress, a derived quantity~\cite{landau2012theory}.

Solids with prestress, nevertheless, exhibit elastic behaviour under small perturbations~\cite{geng2001footprints,otto2003anisotropy,schuh2003atomistic,maloney2006amorphous}, with rigidity emerging from a disordered network of interparticle contacts that enforce local mechanical equilibrium. Jammed solids, composed of non-Brownian particles with purely repulsive interactions, are an extreme example since they acquire rigidity only when an externally applied stress exceeds a critical threshold~\cite{cates1998jamming,liu1998jamming,bi2011jamming,pica2009jamming,o2002random,o2003jamming}. 
This mechanism underlies the physics of jamming transition, a unifying concept of rigidity in disordered materials such as foams, emulsions, and granular media~\cite{liu1998jamming,trappe2001jamming,gopal2003relaxing,chaudhuri2007universal,bi2011jamming}. In Jammed solids, therefore, the mechanical response {\it emerges} from the frozen-in internal stresses.

%%%%%%%%%%%%%%%%%%%%%%%%%%%%%%%%%%%%%%%%%%%%%%%%%%%%%%%%
\begin{figure*}
\centering
  \includegraphics[width=2 \columnwidth]{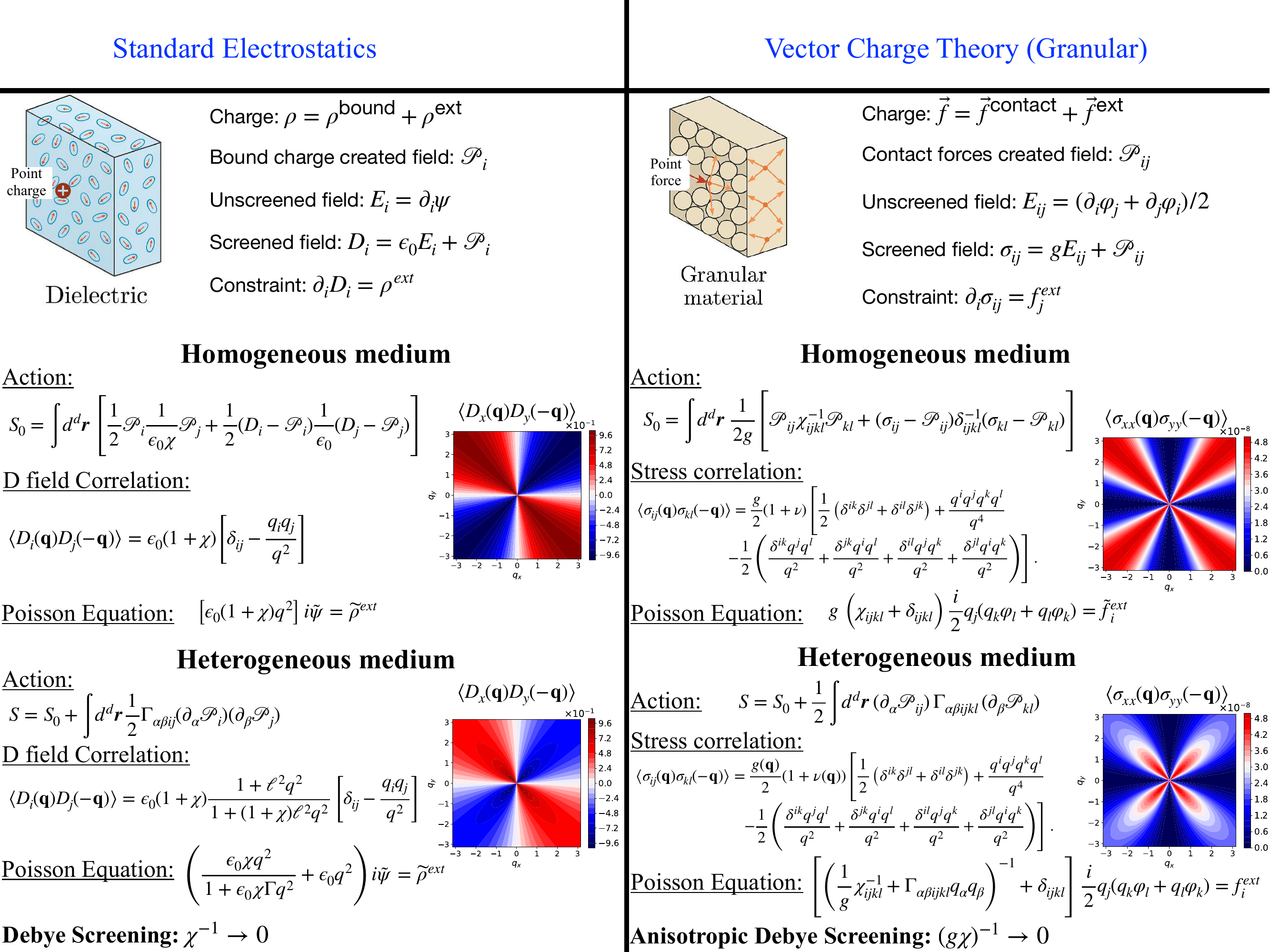}
  \caption{Analogy between dielectric response in standard electromagnetism (EM) and the mechanical response of jammed solids described by VCTG. In both frameworks, the screened fields—$\mathbf{D}$ in EM and $\sigma_{ij}$ in VCTG—combine an unscreened field and a polarization contribution: $\mathbf{D} = \epsilon_0 \mathbf{E} + \boldsymbol{{\mathcal{P}}}$ in EM, and $\sigma_{ij} = g E_{ij} + \mathcal{P}_{ij}$ in granular media, where $E_{ij}$ is a compatible tensor electric field and $\mathcal{P}_{ij}$ encodes internal prestress. Polarization fields follow $P_i = \epsilon_0 \chi E_i$ in linear dielectrics and $\mathcal{P}_{ij} = g \chi_{ijkl} E_{kl}$ is its analog in VCTG. The key distinction is that $g$ depends on the macrostate of the solid and vanishes near unjamming. In heterogeneous media, spatial fluctuations in the polarisation field incur an energetic cost, which suppresses short-wavelength responses. In the EM, this leads to Debye screening in the limit $\chi^{-1} \to 0$, corresponding to highly polarizable materials. A mechanically analogous screening behaviour arises in the VCTG framework, where the limit $(g \chi)^{-1} \to 0$ plays a similar role.}
  \label{summary_figure}
\end{figure*}
%%%%%%%%%%%%%%%%%%%%%%%%%%%%%%%%%%%%%%%%%%%%%%%%%%%%%%%%

Previously, some of the authors of this paper introduced the Vector Charge Theory of Granular mechanics (VCTG) for amorphous elasticity to explain the emergent elasticity of such prestressed solids~\cite{Jishnu_PRL,Jishnu_PRE}. This was established through a mapping of mechanical equilibrium constraints to a tensor \( U(1) \) Tensor Gauge Theories~\cite{pretko2017generalized} with vector charges. The prestress, generated by contact forces, appears as the analogue of the familiar polarisation field present in a dielectric in standard electromagnetism (EM) as summarised in Fig.~\ref{summary_figure}.

VCTG thus provides a stress-only, field-theoretic description~\cite{Jishnu_PRL, Jishnu_PRE} of {\it emergent elasticity} in disordered solids. The conditions of mechanical equilibrium, force and the torque balance are built into its structure via Gauss's law for a symmetric rank-2 tensor electric field sourced by the vector charge -- the force. The {\it self-averaging} jammed solid corresponds to a dielectric of the tensor {\it electromagnetism} where the stress tensor $\sigma_{ij}$ corresponds to the electric displacement field $D_{ij}$. Forces act as vector charges, $\rho_j=\rho_j^{\rm ext}+\rho_j^{\rm bound}$, with $\rho_j^{\rm ext}$ corresponding to external body force and $\rho_j^{\rm bound}$ being the internal contact forces generated in response to the boundary and body forces. Analogous to standard dielectrics~\cite{jackson2021classical}, the vector bound charges induce a tensor polarization field, $\mathcal{P}_{ij}$~\cite{Jishnu_PRL,Jishnu_PRE}. Mechanical response of the jammed solid is characterised by a fourth-rank tensor $\Lambda^{-1}_{ijkl}$, the analogue of a dielectric permittivity, relating stress to the unscreened field via $\sigma_{ij} = \Lambda^{-1}_{ijkl} E_{kl}$. The symmetric field $E_{ij}$, derived from a vector, electrostatic potential, captures the compatible, irrotational response in the absence of internal constraints, while $\sigma_{ij}$ incorporates screening from boundary and body forces~\cite{Jishnu_PRE}. VCTG can also accommodate a Debye-type screening with an anisotropic screening length~\cite{Jishnu_PRE,lemaitre2021anomalous,Kumar:2022aa,fu2025long}.

VCTG successfully predicts long-range anisotropic stress correlations that decay as a power law, $1/r^d$, in $d$ dimensions~\cite{lois2009stress,lemaitre2017inherent,lemaitre2018stress,degiuli2018field,tong2020emergent}, with a non-trivial angular form-factor that gives rise to {\it pinch-point} singularities in Fourier space -- a hallmark of stress correlations in granular systems~\cite{henkes2009statistical,mcnamara2016eshelby,degiuli2018field,wang2020connecting}. Within the VCTG framework, these singularities emerge naturally as a consequence of Gauss's law constraint on the stress tensor. These predictions have been tested in granular solids at high pressures and in gels near the rigidity transition, as well as near-crystalline systems~\cite{vinutha2023stress,countryman2025pinch,Jishnu_PRL,Jishnu_PRE,maharana2024universal,maharana2024stress}.

In this paper, we investigate three mechanical aspects of jammed solids: (i) stress-stress correlations across a range of pressures, approaching unjamming, (ii) the link between displacements of grains in response to external forces, and the potentials of VCTG, and (iii) the role played by disorder averaging in describing the jammed phase, especially close to the unjamming transition. Each of these aspects focuses on important features of the elasticity of jammed solids: (i) addresses the question of whether or not dielectric breakdown precedes unjamming, (ii) resolves the mapping between the electrostatic potential in VCTG and measurements of displacements in disordered, prestressed solids, and (iii) discusses the important role of disorder averaging, particularly close to unjamming where self-averaging is known to fail~\cite{o2003jamming}.

Based on numerical observations of stress-stress correlations, we motivate a natural extension of the VCTG framework, and then numerically examine its consequences. The extended theoretical framework introduces a scale-dependent mechanical response, analogous to a scale-dependent susceptibility in dielectrics. As in the dielectric theory of electrostatics, the extended VCTG framework identifies natural conditions under which the dielectric framework could fail, potentially giving rise to Debye-type screening. However, our analysis reveals no evidence for such behavior in frictionless jammed solids.

\begin{figure}
  \includegraphics[width=0.42\textwidth]{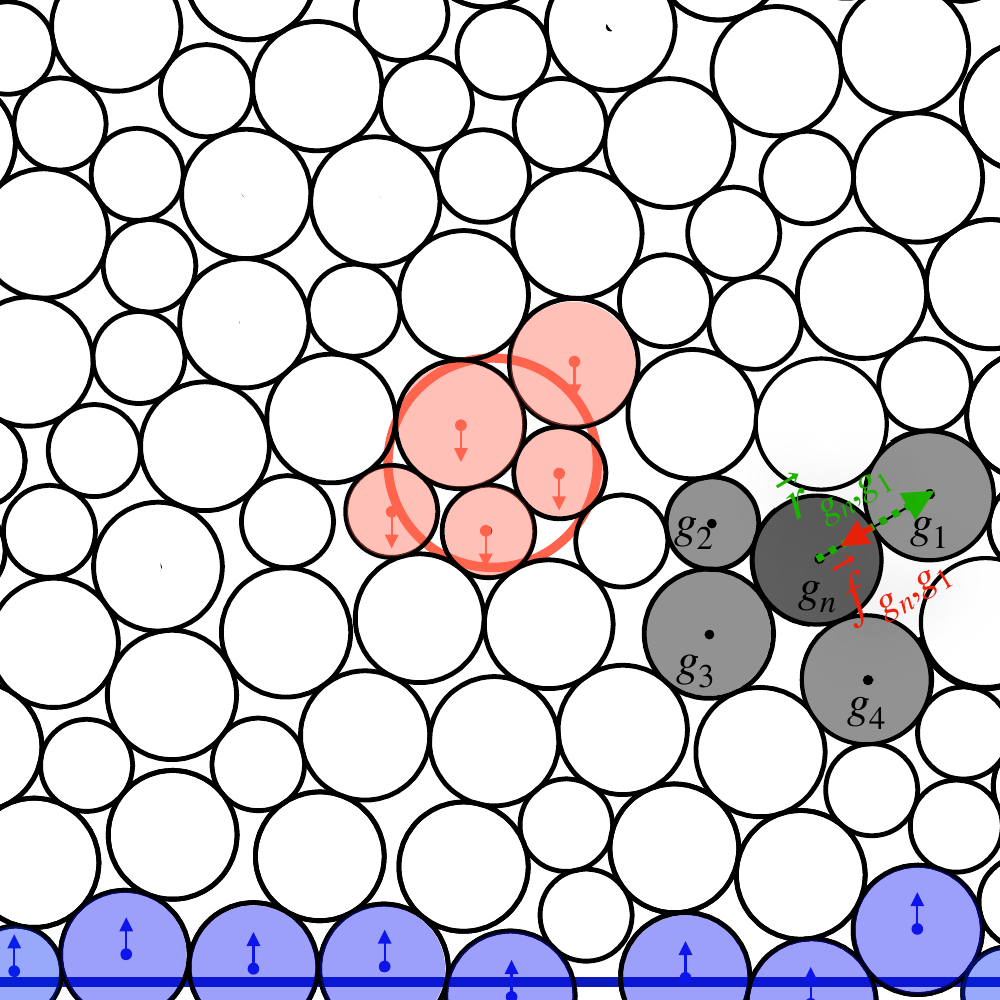}
  \caption{Schematic illustration of the stress tensor convention and external forcing geometry used in this study. The grain-level force-moment tensor at grain $g_{n}$ is defined as $\boldsymbol{\sigma}_{g_{n}} = \sum_m {\bf f}_{g_n,g_m} \otimes {\bf r}_{g_n,g_m}$, where ${\bf f}_{g_n,g_m}$ is the contact force exerted on grain $g_n$ by a neighboring grain $g_m$, and ${\bf r}_{g_n,g_m}$ is the vector connecting the centers of grains $g_n$ and $g_m$. The external force consists of a uniform distribution of force in the negative $y$-direction applied over a circular region centered at the origin over the grains shaded in orange, balanced by a line force in the positive $y$-direction applied at $y = -L/2 + 5$ applied on grains shaded in blue.}
  \label{fig_schematic_force_geometry}
\end{figure}

In addition to the stress response, we compute the displacement of the grains generated by localized deforming forces (Fig.~\ref{fig_schematic_force_geometry}). Stress fields are central to VCTG~\cite{ball2002stress,henkes2005jamming,ramola2017stress}, though they are experimentally accessible only in model systems such as photoelastic disks~\cite{majmudar2005contact,Behringer:2019aa}.
 
In contrast, displacements relative to an unperturbed prestressed configuration can be directly measured in particle-tracking and imaging experiments on granular media, biological tissues and other soft solids, they are widely used in theoretical studies of amorphous elasticity~\cite{lemaitre2021anomalous,mondal2022experimental}. Establishing the link between displacement response and the stress-based field-theoretic formulation of VCTG is therefore essential and can offer a direct route to test the prediction of VCTG across diverse systems.

%%%%%%%%%%%%%%%%%%%%%%%%%%%%%%%%%%%%%%%%%%%%%%%%%%%%%%%%

%%%%%%%%%%%%%%%%%%%%%%%%%%%%%%%%%%%%%%%%%%%%%%%%%%%%%%%%

Finally, we emphasize the crucial role of disorder averaging in revealing the emergent elasticity of amorphous solids. Near the unjamming transition, individual configurations exhibit strong sample-to-sample fluctuations and may undergo irreversible rearrangements in response to even small perturbations, resulting in plastic behavior. However, our numerical results show that ensemble averaging over such configurations—at fixed macroscopic conditions—smooths out these fluctuations and restores a dielectric-like elastic response consistent with VCTG predictions, even up to the unjamming point indicating that the weight of such plastic events averages out within our numerical protocol. This suggests that localized plastic events are statistical in nature and do not define a distinct ``prejammed'' phase in a self-averaging system. Alternatively, restricting attention to those configurations that remain elastic under local perturbations also yields representative responses in line with the VCTG framework. Thus, while it has been proposed that plasticity induces a finite screening length characteristic of a prejammed phase~\cite{lemaitre2021anomalous,fu2025long}, our findings show that the disorder-averaged response of frictionless jammed solids remains well described by the {\it dielectric} within the linear VCTG theory. However, near unjamming the response exhibits pressure-dependent nonlinear deviations, while clear structural changes appear only when the applied force exceeds the confining pressure, a criterion that also governs the response deep in the jammed phase. This effect is distinct from what would be expected from Debye-like screening, which introduces a screening length {\it within} linear response. This observation indicates the need for non-linear corrections to the dielectric formalism of VCTG, and we propose an approach in Appendix~\ref{appen_non_linear}.

The rest of this paper is organized as follows. Section~\ref{sec_correlations} presents numerical results on pressure correlations in jammed soft-sphere packings, highlighting deviations from scale-independent elasticity at short length scales. Section~\ref{sec_vctg_modified} extends the VCTG framework by introducing polarization gradient terms that incorporate a finite length scale while preserving the long-wavelength behavior of the VCTG. Section~\ref{sec_response} compares the predictions of this modified theory with numerical measurements of the mechanical response to localized perturbations, provides a particle-level interpretation of the gauge potential, $\boldsymbol{\varphi}$, as relative displacements, and investigates the pressure dependence of the emergent elastic moduli. Finally, Section~\ref{sec_conclusion} summarizes our findings and outlines possible directions for future research. The Appendices provide additional details: Appendices~\ref{appen_VCTG_DD_corr} and~\ref{sec_averaging} cover technical aspects of the theoretical calculations and numerical analysis, Appendix~\ref{appen_non_linear} discusses the nature of the nonlinear corrections, and Appendix~\ref{Appen_EM} presents the calculation of correlations and response for standard electrostatics.

%%%%%%%%%%%%%%%%%%%%%%%%%%

\section{Pressure Correlations}
\label{sec_correlations}

A key prediction of the original formulation of VCTG~\cite{Jishnu_PRL,Jishnu_PRE} is that, deep inside the jammed phase, different components of the stress tensor exhibit anisotropic correlations. In contrast, correlations of the pressure, the trace of the stress tensor, are isotropic and scale-independent at long wavelengths. In two dimensions, the VCTG calculations~\cite{Jishnu_PRL,Jishnu_PRE} yield the Fourier-space correlation function of local pressure fluctuations,  
\begin{equation}
 \langle P(\mathbf{q}) P(-\mathbf{q}) \rangle 
 = \frac{g (1 + \nu)}{2} 
 \equiv K_{\mathrm{2D}},
\label{eq_P_corr_K2D} 
\end{equation}  
where $P(\mathbf{q})$ denotes the Fourier components of the local pressure fluctuations (throughout this paper, we denote spatial fluctuations of the local pressure field by $P$, while $p$ refers to the global pressure of the system) with the average, $\langle\cdots\rangle$, taken over independent packings at fixed pressure, $p$. Here, $g$, and $\nu$, are coupling constants of VCTG (see Fig.~\ref{summary_figure}), and characterize the mechanical response of the jammed solid. As will be discussed in Section~\ref{sec_response}, these coupling constants play roles analogous to the Lam\'e coefficients in classical elasticity theory, with $g$ being the shear modulus and $\nu$ the Poisson ratio. Thus, pressure correlations bear the signature of the emergent elasticity via, $g$, and $\nu$.

We start by numerically examining the validity of the prediction in Eq.~\eqref{eq_P_corr_K2D} as a system approaches unjamming from the jammed side. We simulate two-dimensional jammed packings of a $50:50$ bidisperse mixture of frictionless disks with a diameter ratio of $1:1.4$, under periodic boundary conditions. Particles interact via short-range repulsive potential,
\begin{equation}
v(r_{ij}) = \frac{\epsilon}{\alpha} \left( 1 - \frac{r_{ij}}{a_{ij}} \right)^\alpha, \label{eqn_potn}
\end{equation}
where $r_{ij}$ is the inter-particle distance between particles $i$ and $j$, and $a_{ij} = \frac{a_i + a_j}{2}$ is the sum of their radii. We consider both Harmonic ($\alpha=2$) and Hertzian ($\alpha=\frac{5}{2}$) interactions. Energies are measured in units of $\epsilon$ and lengths in units of the smaller particle diameter $a_0$. We set $\epsilon = 1$ and $a_0 = 1$ for convenience. Jammed configurations at various pressures $p$ are generated via conjugate gradient energy minimization, following~\cite{o2002random}.

%%%%%%%%%%%%%%%%%%%%%%%%%%%%%%%%%%%%%%%%%%%%%%%%%%%%%%%%%%%%%%%%%%%%%
\begin{figure}[t!]
  \includegraphics[width=0.475\textwidth]{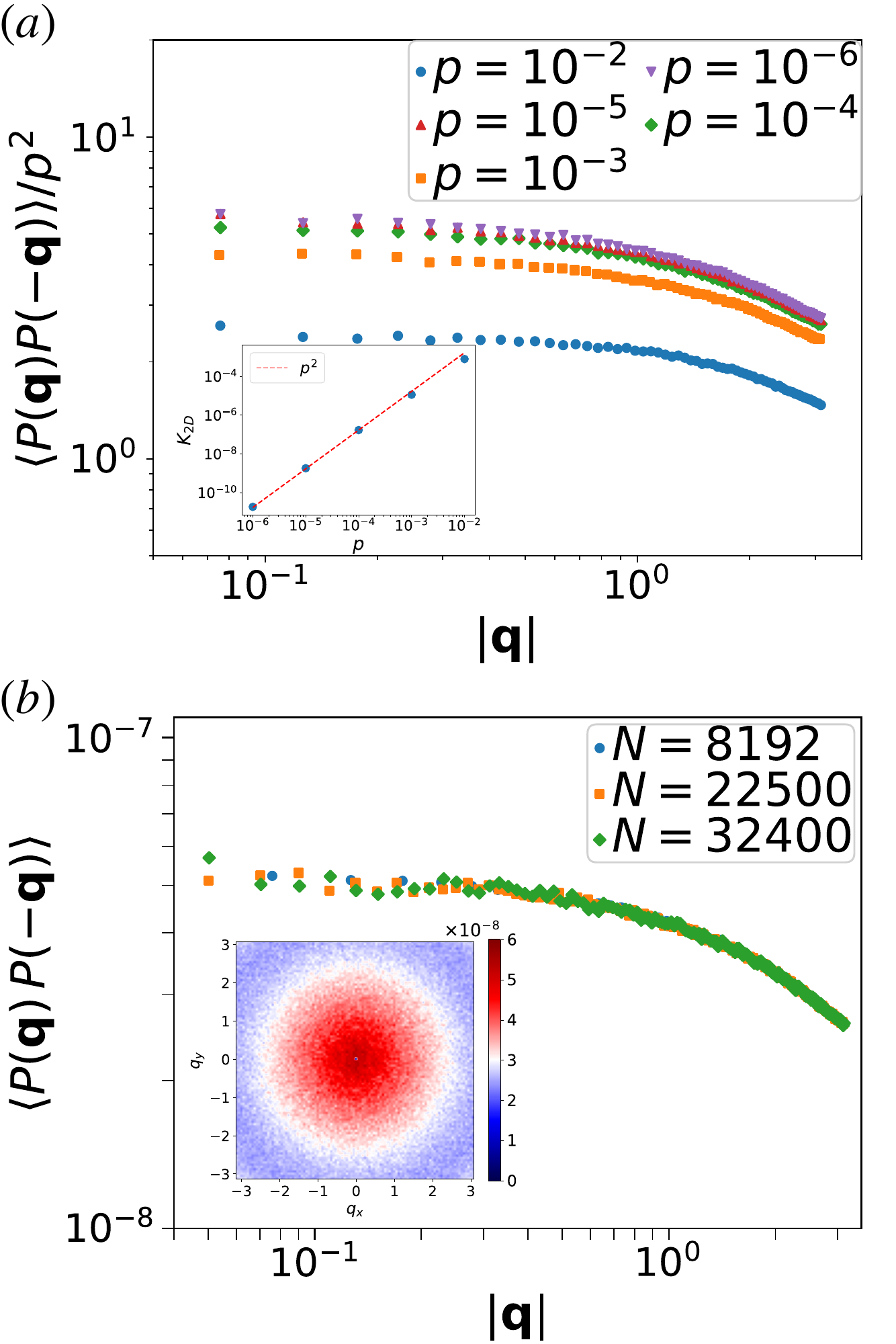}
  \caption{Fourier-space pressure correlation function, $\langle P(\boldsymbol{q}) P(-\boldsymbol{q}) \rangle$, for harmonic packings of $N = 8192$ particles at pressures approaching the unjamming transition. (a) Correlation normalized by $p^{2}$, exhibiting a plateau whose height defines $K_{2D}$ [Eq.~\eqref{eq_P_corr_K2D}]. The data collapse at low $p$ and long-wavelength behavior reflects the scale-invariant correlations at large length scales. At large $|\boldsymbol{q}|$, the correlation decays in a pressure-independent manner, indicating universal suppression of short-wavelength pressure fluctuations. {\it Inset} shows that $K_{2D}$ scales as $p^2$ near unjamming. (b) Pressure correlation as a function of $|\boldsymbol{q}|$ at fixed $p = 10^{-4}$ for different system sizes, highlighting finite-size effects and convergence at both long and short wavelengths. Inset: Two-dimensional $\boldsymbol{q}$-space map showing isotropic correlations, with a crossover from a flat, long-wavelength regime to a short-wavelength decay. 
  }
  \label{fig_pressure_corr}
\end{figure}
%%%%%%%%%%%%%%%%%%%%%%%%%%%%%%%%%%%%%%%%%%%%%%%%%%%%%%%%%%%%%%%%%%%%%

The grain-level force-moment tensor is obtained as
\begin{equation}
    \boldsymbol{\sigma}_{g_n} = \sum_m {\bf f}_{g_n,g_m} \otimes \boldsymbol{r}_{g_n,g_m},
    \label{eq_grainforcemoment}
\end{equation}
where ${\bf f}_{g_n,g_m}$ is the contact force exerted on grain $g_n$ by a neighboring grain $g_m$, and $\boldsymbol{r}_{g_n,g_m}$ is the vector connecting their centers (Fig.~\ref{fig_schematic_force_geometry}). The above force-moment tensor has the dimensions of [energy]. This force-moment tensor, $\boldsymbol{\sigma}_g$, is readily averaged over and has straightforward generalizations to amorphous materials such as glasses, where longer-ranged interactions may be present. We however refer to this as stress instead of the force-moment tensor for brevity.

For a given configuration, the spatially averaged stress is defined as
\begin{equation}
\boldsymbol{\bar{\sigma}} = \frac{1}{N} \sum_{g_n=1}^{N} \boldsymbol{\sigma}_{g_n},
\end{equation}
where $N=L^d \phi$ is the total number of grains in the system, and $\phi$ is the packing fraction. Local stress fluctuations relative to this average are then defined at the center of each grain, $r_{g_n}$, as
\begin{equation}
\Delta\sigma_{ij}(\boldsymbol{r}_{g_n}) = \sigma_{ij}(\boldsymbol{r}_{g_n}) - \bar{\sigma}_{ij}.
\label{eq_sigmafluc}
\end{equation}

The discrete Fourier transform of the stress fluctuations is given by
\begin{equation}
 \Delta\sigma_{ij}(\boldsymbol{q})= \frac{1}{\sqrt{N}}\sum_{g_n=1}^N\, e^{i \boldsymbol{q}\cdot\boldsymbol{r}_{g_n}}\, \Delta\sigma_{ij}(\boldsymbol{r}_{g_n}). \label{eq_Fourier_stress_fluc}
\end{equation}

Note that in a periodic box of linear size, $L$, the spacing of the Fourier modes is given by $\Delta q = 2\pi/L$ in each direction. Finally, the Fourier-space stress correlations are computed as  
\begin{equation}
 C_{ijkl}(\boldsymbol{q} ) = \langle\Delta\sigma_{ij}(\boldsymbol{q})\Delta\sigma_{kl}(-\boldsymbol{q})\rangle.
 \label{eq_stress_corr}
\end{equation} 
 
Fig.~\ref{fig_pressure_corr}(a) shows $\langle P(\boldsymbol{q}) P(-\boldsymbol{q}) \rangle$ as a function of wavevector magnitude $|{\bf q}|$ for harmonic packings of $N = 8192$ particles across a range of pressures, approaching the unjamming threshold. At small $|{\bf q}|$, corresponding to long wavelengths, the correlation displays a plateau, consistent with the scale-invariant predictions of VCTG [Eq.~\eqref{eq_P_corr_K2D}]. The \textit{inset} demonstrates that the plateau height, $K_{\text{2D}}$, scales as $p^2$ near unjamming, consistent with the pressure dependence of effective moduli in two dimensions~\cite{henkes2009statistical,chakraborty2024long} such that in the main panel, where we plot $\langle P(\boldsymbol{q}) P(-\boldsymbol{q}) \rangle / p^2$ the data collapses across pressures in the low-$p$ regime and small $|\boldsymbol{q}|$.

Importantly, at large $|\boldsymbol{q}|$, the correlation systematically decays, indicating a suppression of pressure fluctuations at short distances. Notably, the shape of this decay is independent of pressure near unjamming, pointing to a universal deviation from scale invariance at high wavevectors. Fig.~\ref{fig_pressure_corr}(b) explores finite-size effects by plotting $\langle P(\boldsymbol{q}) P(-\boldsymbol{q}) \rangle$ versus $|\boldsymbol{q}|$ for various system sizes at fixed pressure $p = 10^{-4}$, confirming the scale-free plateau at smaller $|\boldsymbol{q}|$ and the decay at shorter wavelengths. The inset of Fig.~\ref{fig_pressure_corr}(b) displays the two-dimensional correlation map, revealing the isotropic nature of pressure fluctuations across all wavevectors, and clearly illustrating the crossover from a flat, long-wavelength plateau to a decay regime at higher $|{\bf q}|$. 

This short-distance decay reflects the limitations of a purely continuum description and implies the presence of a finite length scale below which the Gaussian theory of VCTG must be modified. In real space, this corresponds to a crossover beyond which the power-law decay of stress correlations emerges. {We quantify this scale later and find it to be approximately two grain diameters}. These observations naturally lead to the question: Can VCTG, which captures long-range, scale-invariant features, be extended to account for the finite-wavelength suppression of stress fluctuations?

In the following, we modify the VCTG theory~\cite{Jishnu_PRL,Jishnu_PRE} to incorporate an intrinsic microscopic length scale. This extension preserves the scale-invariant features at long distances while regularizing short-range behavior consistent with our numerical observations.

\section{Modified Field Theory and additional Lengthscale}\label{sec_vctg_modified}

Within the VCTG, 
stress–stress correlations and linear response functions are computed from the partition function in the presence of an external body force $f_j^{\rm ext}$~\cite{Jishnu_PRE},  
\begin{equation}
\mathcal{Z}[f_j^{\rm ext}] = \int [\mathcal{D}\mathcal{P}][\mathcal{D}\sigma] \, 
\delta(\partial_i \sigma_{ij}-f_j^{\rm ext}) \, e^{-\mathcal{S}_0},
\end{equation}
where the delta function enforces Gauss’s law. To lowest order, the Gaussian action, $\mathcal{S}_0$, is given in Fig.~\ref{summary_figure}.

%%%%%%%%%%%%%%%%%%%%%%%%%%%%%%%%%%%%%%%%%%%%%%%%%%%%%%%%
\begin{figure}%[t!]
  \includegraphics[width=0.45\textwidth]{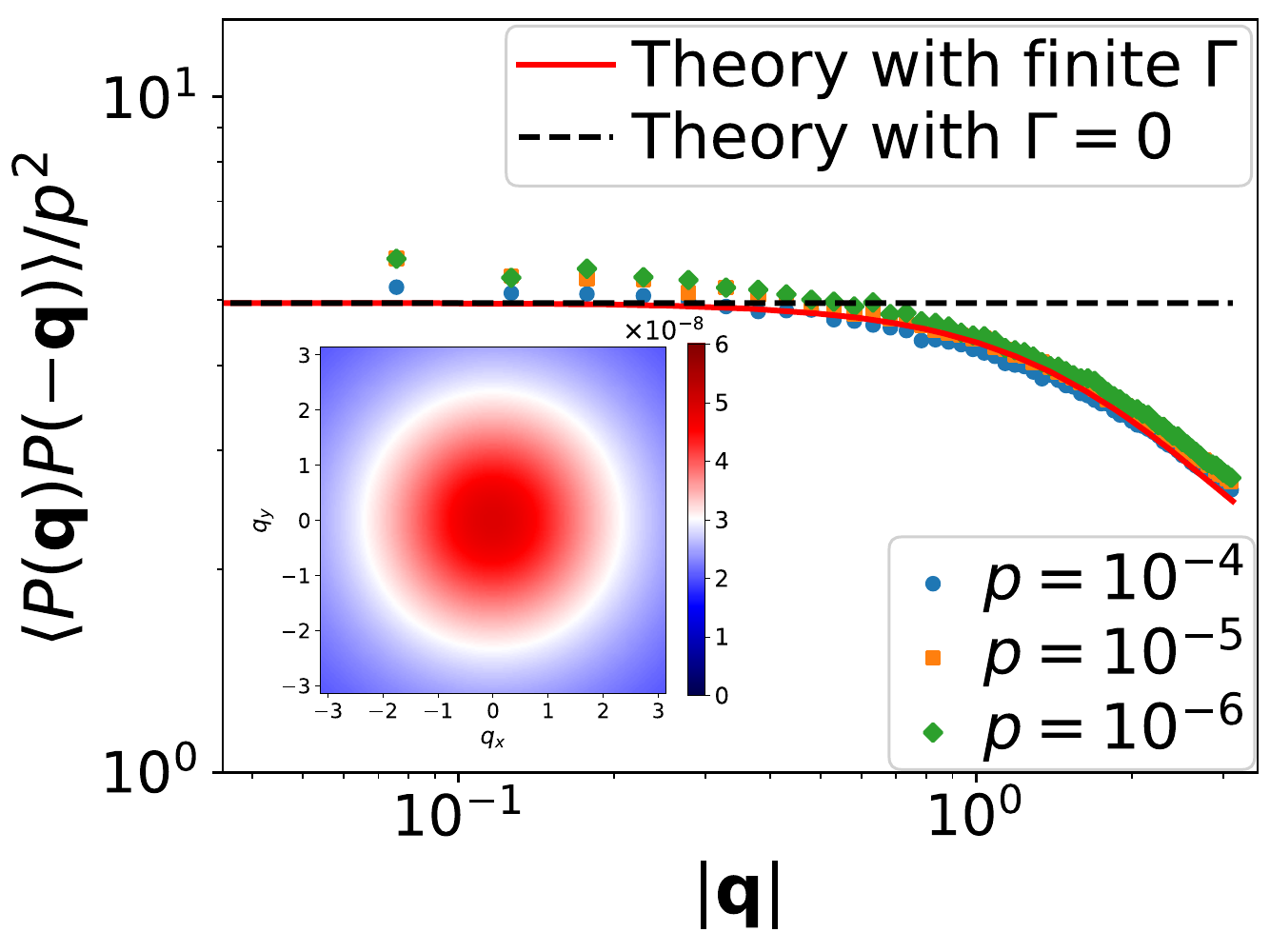}
  \caption{Pressure correlation near unjamming compared with predictions of the extended VCTG including polarization gradient terms [Eq.~\eqref{eq_P_corr_main}]. Simulation data for $p = 10^{-4}, 10^{-5}, 10^{-6}$ are shown scaled by $p^{2}$. The theoretical curve is fitted at $p = 10^{-4}$ with $\nu = 0.95$, with $g$ chosen to match the low-$|{\bf q}|$ plateau, and is scaled by the same factor for direct comparison. A finite polarization stiffness, $g\Gamma = 0.062$, reproduces the large-$|{\bf q}|$ decay and captures the crossover.}
  \label{fig_p_corr_fits}
\end{figure}
%%%%%%%%%%%%%%%%%%%%%%%%%%%%%%%%%%%%%%%%%%%%%%%%%%%%%%%%

%%%%%%%%%%%%%%%%%%%%%%%%%%%%%%%%%%%%%%%%%%%%%%%%%%%%%%%%
\begin{figure*}[t!]
\centering
  \includegraphics[width=1.9\columnwidth]{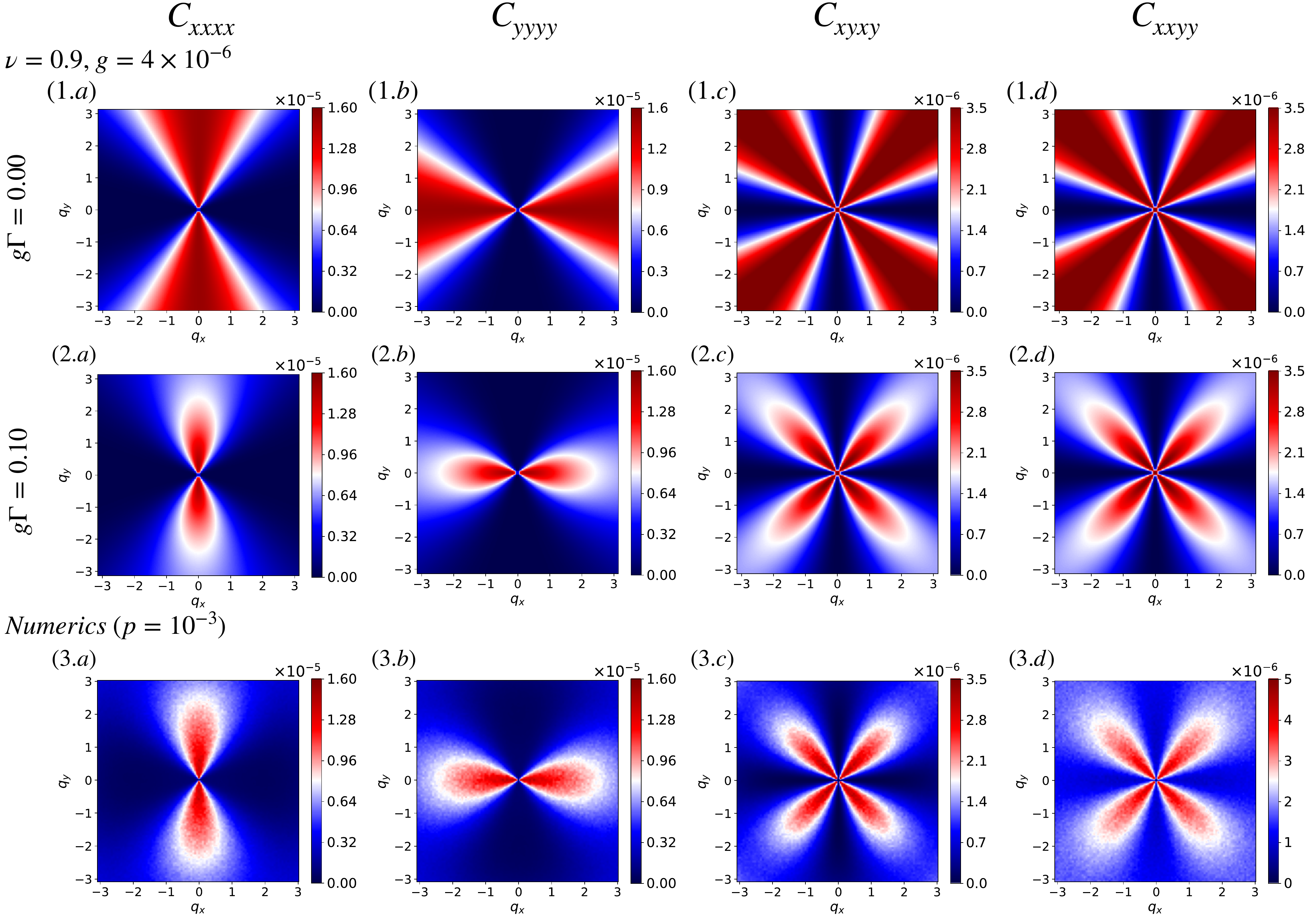}
  \caption{Fourier-space stress correlation functions, $\langle \sigma_{ij}(\mathbf{q}) \sigma_{kl}(-\mathbf{q}) \rangle$, obtained from Eq.~\eqref{eq_stresscorr} with an isotropic polarization stiffness tensor. The columns display the four components $C_{xxxx}$, $C_{yyyy}$, $C_{xyxy}$, and $C_{xxyy}$. The first two rows show the effect of increasing $\Gamma$ at fixed $\nu = 0.9$ and $g = 4\times 10^{-6}$, parameter values chosen consistently with the stress-response analysis and with correlations measured at pressure $p=10^{-3}$. For $\Gamma=0$ (top row), the correlations are independent of $\lvert \mathbf{q} \rvert$ and anisotropic. Finite $\Gamma$ progressively suppresses large-$|\mathbf{q}|$ fluctuations, and at sufficiently large $\Gamma$ the correlations are strongly attenuated except near small $|\mathbf{q}|$, where anisotropic pinch-point singularities persist (second row). The third row shows corresponding correlations measured in simulations of jammed packings at pressure $p=10^{-3}$.
  }
  \label{fig_corr_fixed_mu}
\end{figure*}
%%%%%%%%%%%%%%%%%%%%%%%%%%%%%%%%%%%%%%%%%%%%%%%%%%%%%%%%

To capture the short-lengthscale deviations from the scale-independent VCTG framework, as evident in the pressure correlation obtained from simulations of jammed solids (Fig.~\ref{fig_pressure_corr}), we extend the theory by incorporating the leading symmetry-allowed gradient terms to obtain the extended action 
\begin{align} 
\mathcal{S}=\mathcal{S}_0+\mathcal{S}_1, \label{eq_action_eVCTG} 
\end{align} 
with the correction term 
\begin{align} 
\mathcal{S}_1=\frac{1}{2}\int d^d \boldsymbol{r} \, ( \partial_\alpha \mathcal{P}_{ij}) \, \Gamma_{\alpha \beta ijkl} \, (\partial_\beta \mathcal{P}_{kl}). \label{eq_augmented_action} 
\end{align} 
Here, $\Gamma_{\alpha\beta ijkl}$ is a sixth-rank tensor that encodes the stiffness associated with gradients of the polarization field $\mathcal{P}_{ij}$. This correction penalizes local inhomogeneities in the internal stress organization. This formalism is analogous to the treatment of gradient terms in dielectric theories of standard electromagnetism, where spatial variations in the polarization field $\boldsymbol{\mathcal{P}}$ contribute an energy density $\propto |\nabla \boldsymbol{\mathcal{P}}|^2$. In standard dielectrics, such terms arise from the electrostatic energy of bound charges, which penalize non-uniform polarization. The analogy is illustrated in Fig.~\ref{summary_figure} and discussed further in Appendix~\ref{Appen_EM}.

Implementing the constraint of Gauss's law via a Lagrange multiplier field, $\boldsymbol{\varphi}$, we obtain %the partition function 
\begin{equation}
\mathcal{Z}[f_j^{\rm ext}] = \int [\mathcal{D}\mathcal{P}][\mathcal{D}\sigma][\mathcal{D}\varphi] ~ e^{-\tilde{\mathcal{S}}}, \label{eq_partition_fn}
\end{equation}
with the effective action  
\begin{align}
\tilde{\mathcal{S}} = &\int d^d \mathbf{r} \left[ \frac{1}{2g} \Big\{ \mathcal{P}_{ij} \chi^{-1}_{ijkl} \mathcal{P}_{kl} + (\sigma_{ij} - \mathcal{P}_{ij}) \delta^{-1}_{ijkl} (\sigma_{kl} - \mathcal{P}_{kl}) \Big\}\right.\nonumber\\
&~~~~~~+ \left.\frac{1}{2} (\partial_\alpha \mathcal{P}_{ij}) \, \Gamma_{\alpha \beta ijkl} (\partial_\beta \mathcal{P}_{kl}) + i \sigma_{ij} J_{ij} + i \varphi_i f_i^{\rm ext} \right].\label{MVCTG_eq_Constraint}
\end{align}

Here, the coupling constant $g$, has dimensions of $\text{[energy]}^2 \text{[length]}^d$, since the stress, $\sigma_{ij}$, and the polarization, $\mathcal{P}_{ij}$, are actually force-moment tensors, which have dimensions of $\text{[energy]}$. $\mathcal{J}_{ij}$ is an auxiliary source field that couples linearly to the total field $\sigma_{ij}$ enabling the computation of stress-stress correlations via functional derivatives of $\log \mathcal{Z}[J]$ and $J_{ij}= \mathcal{J}_{ij} + \tfrac{1}{2}(\partial_i \varphi_j + \partial_j \varphi_i)$. The detailed calculation is presented in Appendix~\ref{appen_VCTG_DD_corr}.

To incorporate the consequences of polarization gradients, we consider the simplest isotropic form of the polarization stiffness tensor introduced in Eq.~\eqref{eq_augmented_action}, 
\begin{equation}
 \Gamma_{\alpha\beta ijkl} = \delta_{\alpha\beta} \,\frac{\Gamma}{2}\, \left( \delta_{ik}\delta_{jl} + \delta_{il}\delta_{jk} \right)\label{eq_stiffness_tensor}
 \end{equation} 
with $\Gamma$ having dimensions of [length]$^{2-d}$[energy]$^{-2}$. This choice assigns equal stiffness to all components of $\mathcal{P}_{ij}$ and penalizes spatial variations isotropically, thereby preserving rotational invariance, ensuring that the trace of the stress tensor exhibits isotropic correlations. The resulting pressure–pressure correlation function reads 
\begin{equation}
\langle P(\boldsymbol{q}) P(-\boldsymbol{q}) \rangle = \frac{g (2 + g \Gamma q^2) \left[ 2 (1 + \nu) + g \Gamma q^2 (1 + 3 \nu) \right]} {8 + 4 g  \Gamma q^2 \left[ 3 + 5 \nu + g \Gamma q^2 (1 + 3 \nu) \right]}, \label{eq_P_corr_main} 
\end{equation}
which reduces in the long-wavelength limit ($|{\bf q}| \to 0$) to a scale-independent plateau given by Eq.~\eqref{eq_P_corr_K2D}. The parameter $\Gamma$ introduces a characteristic length scale, 
\begin{equation}
\ell = \sqrt{g \Gamma}, 
\label{eq_scale}
\end{equation} 
For wavevectors $|{\bf q}| \gtrsim |{\bf q}|^* \sim 1/\ell$, stress correlations deviate from their scale-free form. The length-scale, $l$, should be compared with the standard electrostatic screening length 
$\lambda =\sqrt{\epsilon_0 \Gamma}$ presented in Appendix~\ref{Appen_EM}. A crucial point of difference between the VCTG screening length and the electrostatic one is that $\epsilon_0$ is a fundamental constant, whereas $g$ depends on the proximity to rigidity and goes to zero at the unjamming transition.

Fig.~\ref{fig_p_corr_fits} compares the scaled simulation data with Eq.~\eqref{eq_P_corr_main}. To compare to numerical results, we fix $\nu = 0.95$, consistent with the stress response analysis shown in Section \ref{sec_response}. The coupling constant, $g$, is chosen to match the $|{\bf q}| \to 0$ plateau at $p = 10^{-4}$, and the theoretical curve is scaled by $p^2$ for comparison. A value $g\Gamma = 0.062$ captures the observed decay, yielding quantitative agreement with the data. Importantly, the crossover wavevector $|{\bf q}|^*$ is finite and pressure-independent, implying a microscopic screening length $\lvert \mathbf{r}^*\rvert \sim 2\pi \ell$, approximately two grain diameters. Beyond this scale, stress correlations revert to the continuum $1/r^d$ decay. Other components of the stress–stress correlations exhibit anisotropy [Eq.~\eqref{eq_stresscorr}], with wavevector dependence entering through the amplitude $\tilde{K}_{2D}(\mathbf{q})$. Representative results are shown in Fig.~\ref{fig_corr_fixed_mu} for $\nu = 0.9$, $g = 4 \times 10^{-6}$, chosen to match simulation results at $p = 10^{-3}$. When $\Gamma = 0$, the theory reduces to the scale-free VCTG with pinch points. Finite $\Gamma$ suppresses large-$|\mathbf{q}|$ fluctuations, leading to a crossover to short-range behavior while preserving pinch-point singularities at small $|\mathbf{q}|$ [Eq.~\eqref{eq_stresscorr}]. The bottom row illustrates corresponding correlations measured in jammed packings at $p=10^{-3}$.

\section{Response to external forces}\label{sec_response}

We now turn to analyzing the effects of the polarization gradient terms [Eq.~\eqref{eq_augmented_action}] on the mechanical response of jammed solids to external perturbations, as we tune $p$ towards unjamming. For this we use a protocol, similar to that of Ref.~\cite{Jishnu_PRE}, and illustrated in Fig.~\ref{fig_schematic_force_geometry}.

To probe the response to external forces in real space, we compute coarse-grained stress and displacement fields. The coarse-grained stress field is given by
 \begin{equation}
 \boldsymbol{\sigma}(\boldsymbol{r}) = \frac{1}{N_\Omega} \sum_{g_n \in \Omega} \boldsymbol{\sigma}_{g_n}, \label{eq_stressnorm}
 \end{equation} 
where $N_{\Omega}$ is the number of grains in the coarse-graining region $\Omega$.

Similarly, grain displacements ${\bf u}_{g_i}$, from an initial reference configuration in static equilibrium are coarse-grained using the same procedure to obtain the displacement field

\begin{equation}
{\bf u}(\boldsymbol{r}) = \frac{1}{N_\Omega}\sum_{i \in \Omega} {\bf u}_{g_i}.
\end{equation}

We note that within a static configuration, $N_{\Omega}$, in each coarse-graining box can vary, which can be an additional source of fluctuations in the coarse-grained fields. While coarse-graining is convenient for visualizing the spatial profile of stresses and displacements, stress correlations can be evaluated directly in Fourier space, without coarse-graining, as discussed in Section~\ref{sec_correlations}.

The gauge redundancy intrinsic to VCTG, reflects the absence of a unique zero-stress reference state, which results in the ambiguity in defining displacement vectors relative to any arbitrary reference configuration. As a result, displacements from {\it equally valid} reference configurations are {\it unobservable} in configuration averaged correlations and responses~\cite{Jishnu_PRE,Jishnu_PRL}. In practice, however, numerical and experimental studies naturally resolve this issue by measuring displacement differences, analogous to differences in electrostatic potential, $\boldsymbol{\varphi}$. By measuring displacements relative to configurations in the absence of the localised external forces, we can make direct comparisons with the {\it difference} in the electrostatic potential, $\boldsymbol{\varphi}$, although the absolute value of the latter is {\it unobservable} in configuration averaged responses as discussed above. This establishes a direct correspondence between the gauge-theoretic formulation of the response functions and classical elasticity with $\boldsymbol{\varphi}$ playing a role analogous to the relative displacement field $\boldsymbol{u}$, 
albeit without reference to a unique stress-free configuration. However, the mapping of the displacement fields to $\boldsymbol{\varphi}$, also involves a careful analysis of the dimensions of these fields, and we address this in the context of determining the pressure dependence of the elastic moduli.

To compute the responses in the stress and $\boldsymbol{\varphi}$ fields, we perform a saddle-point analysis of the constrained action defined in Eq.~\eqref{MVCTG_eq_Constraint}. At the saddle point of the action with respect to the gauge potential $\boldsymbol{\varphi}$, the Gauss-law constraint yields the condition of mechanical equilibrium,
\begin{equation}
\partial_i \sigma_{ij} = f_j^{\rm ext}.
\end{equation}
Variations of the action with respect to $\sigma_{ij}$ and $\mathcal{P}_{ij}$ respectively yield the two constitutive equations 
\begin{align}
\sigma_{ij} &= \mathcal{P}_{ij} - \frac{i}{2}g \delta_{ijkl}(\partial_k \varphi_l + \partial_l \varphi_k), \label{eq_stress_from_phi} \\
\chi^{-1}_{ijkl} \mathcal{P}_{kl} &- \delta^{-1}_{ijkl}(\sigma_{kl} - \mathcal{P}_{kl}) - g \Gamma_{\alpha\beta ijkl} \partial_\alpha \partial_\beta \mathcal{P}_{kl} = 0. \label{eq_polarization_from_phi}
\end{align}

Substituting Eq.~\eqref{eq_stress_from_phi} into Eq.~\eqref{eq_polarization_from_phi} eliminates $\sigma_{ij}$, which, followed by a Fourier transform gives:
\begin{equation}
\mathcal{P}_{kl}(\boldsymbol{q}) = g\, \tilde{\chi}_{ijkl}(\boldsymbol{q}) \frac{1}{2}(q_i \varphi_j + q_j \varphi_i),\label{eq_modified_P_in_q}
\end{equation}
where 
\begin{align}
    \tilde{\chi}_{ijkl}(\boldsymbol{q})= \left(\chi^{-1}_{ijkl} + g \Gamma_{\alpha\beta ijkl} q_\alpha q_\beta\right)^{-1}
    \label{eq_modified_chi}
\end{align}
is the scale-dependent susceptibility tensor. {Eq.~\eqref{eq_modified_P_in_q} mirrors the constitutive relation of dielectric theory of standard EM with polarization-gradient terms (Appendix \ref{Appen_EM}), where the vector polarization 
couples to the vector electric field through a scale-dependent susceptibility. The corresponding real-space relation for the vector polarization field is given in Eq.~\eqref{eq:modified_P}, which, upon Fourier transformation, reduces to the same underlying form as Eq.~\eqref{eq_modified_P_in_q}, despite its different tensorial character.} 

Substituting back into Eq.~\eqref{eq_stress_from_phi}, we obtain the stress field in Fourier space:
\begin{align}
\sigma_{ij}(\boldsymbol{q}) &= g\, \left( \tilde{\chi}_{ijkl}(\boldsymbol{q}) + \delta_{ijkl} \right) \frac{1}{2}(q_k \varphi_l + q_l \varphi_k) \\
&\equiv  \tilde{\Lambda}^{-1}_{ijkl}(\boldsymbol{q}) \frac{1}{2}(q_k \varphi_l + q_l \varphi_k), \label{eq_sigma_from_phi_saddle}
\end{align}
where $\boldsymbol{\tilde{\Lambda}}^{-1}$ { is defined as} the effective scale-dependent elastic tensor related to $\tilde{\boldsymbol{\chi}}$ (see Eq.~\eqref{eq_modified_lambda_inv}).

In the $\bf{q} \rightarrow 0$ limit, upon disorder averaging, the dielectric permittivity tensor, $\Lambda^{-1}_{ijkl}$, reduces to an isotropic form~\cite{Jishnu_PRE,Jishnu_PRL}. In analogy with isotropic elasticity theory, whose elastic modulus tensor, $K_{ijkl}$, is described by the two Lamé coefficients $(\mu,\nu)$, the shear modulus and the Poisson ratio, we parametrize $\Lambda^{-1}_{ijkl}$ as:  
\begin{align}
\nonumber
\Lambda^{-1}_{ijkl} &= g \big(\delta_{ijkl} + \chi_{ijkl}\big), \\
\chi_{ijkl} &= \tfrac{1}{2}(\delta_{ik}\delta_{jl}+\delta_{il}\delta_{jk}) + \tfrac{2\nu}{1-\nu}\,\delta_{ij}\delta_{kl},\label{eq_elasticity_tensor}
\end{align}
where $\delta_{ijkl}=\tfrac{1}{2}(\delta_{ik}\delta_{jl}+\delta_{il}\delta_{jk})$ is the fourth-rank identity tensor.

Inserting these into the Gauss-law leads to a Poisson equation for the potential :

\begin{widetext}
\begin{equation}
i \left[ \left( (g \chi)^{-1}_{ijkl} + \Gamma_{\alpha\beta ijkl} q_\alpha q_\beta \right)^{-1} + g \delta_{ijkl} \right] \frac{1}{2} q_i (q_k \varphi_l + q_l \varphi_k) = f_j^{\mathrm{ext}}(\boldsymbol{q}). \label{eq:phi_poisson}
\end{equation}
\end{widetext}

Eq.~\eqref{eq:phi_poisson} is equivalent to a generalized Cauchy–Navier equation for an isotropic elastic medium~\cite{ciarlet1994three} 
\begin{equation}
i \left[ \left( \tilde{g}(\boldsymbol{q}) \frac{1 + \tilde{\nu}(\boldsymbol{q})}{1 - \tilde{\nu}(\boldsymbol{q})} \right) q_i q_j + \tilde{g}(\boldsymbol{q})\, q^2 \delta_{ij} \right] \varphi_j(\boldsymbol{q}) = \tilde{f}_i(\boldsymbol{q}).
\end{equation}
with scale-dependent moduli. The ${\bf q}$-dependence introduced by the polarization gradient effects are given by
\begin{equation}
\tilde{g}(\boldsymbol{q}) = \frac{g (2 + g\Gamma q^2)}{2(1 + g\Gamma q^2)},\label{scale_dependent_mu}
\end{equation}
and
\begin{equation}
\tilde{\nu}(\boldsymbol{q}) = \frac{2 \nu}{2 + g \Gamma q^2 \left(3 + 5 \nu + g  \Gamma q^2 (1 + 3 \nu)\right)}\label{scale_dependent_nu}.
\end{equation}
This demonstrates that the polarization stiffness $\Gamma$ introduces a characteristic length scale that modifies the effective elastic response of the system.

The corresponding Green's function $\tilde{c}^{-1}_{ij}(\boldsymbol{q})$, defined via $i \varphi_i = \tilde{c}^{-1}_{ij} f^{\rm ext}_j$, is given by
\begin{equation}
\tilde{c}^{-1}_{ij}(\boldsymbol{q}) = \frac{1}{\tilde{g }(\boldsymbol{q}) q^2} \left( \delta_{ij} - \frac{1 + \tilde{\nu}(\boldsymbol{q})}{2} \frac{q_i q_j}{q^2} \right).\label{eq_Greens_phi}
\end{equation}
From Eq.~\eqref{eq_sigma_from_phi_saddle}, the stress response is
\begin{equation}
\sigma_{ij} = \widetilde{G}_{ijk} f^{\rm ext}_k,
\end{equation}
with Green's function
\begin{equation}
    \widetilde{G}_{ijk}(\boldsymbol{q}) = \frac{1}{q^2} \left[ \tilde{\nu}\, q_k \delta_{ij} + q_i \delta_{kj} + q_j \delta_{ki} - (1 + \tilde{\nu}) \frac{q_i q_j q_k}{q^2} \right]. \label{eq_Greens_stress}
\end{equation}

%%%%%%%%%%%%%%%%%%%%%%%%%%%%%%%%%%%%%%%%%%%%%%%%%%%%%%%%
%%%%%%%%%%%%%%%%%%%%%%%%%%%%%%%%%%%%%%%%%%%%%%%%%%%%%%%%%%%%%%%%%%%%%%%%%%%%%%%%%%%%%%%%%%%%%%%%%%%%%%%%%%%%%%%%%%%%%%%%%%%%
\begin{figure*}
\centering
  \includegraphics[width=1.9\columnwidth]{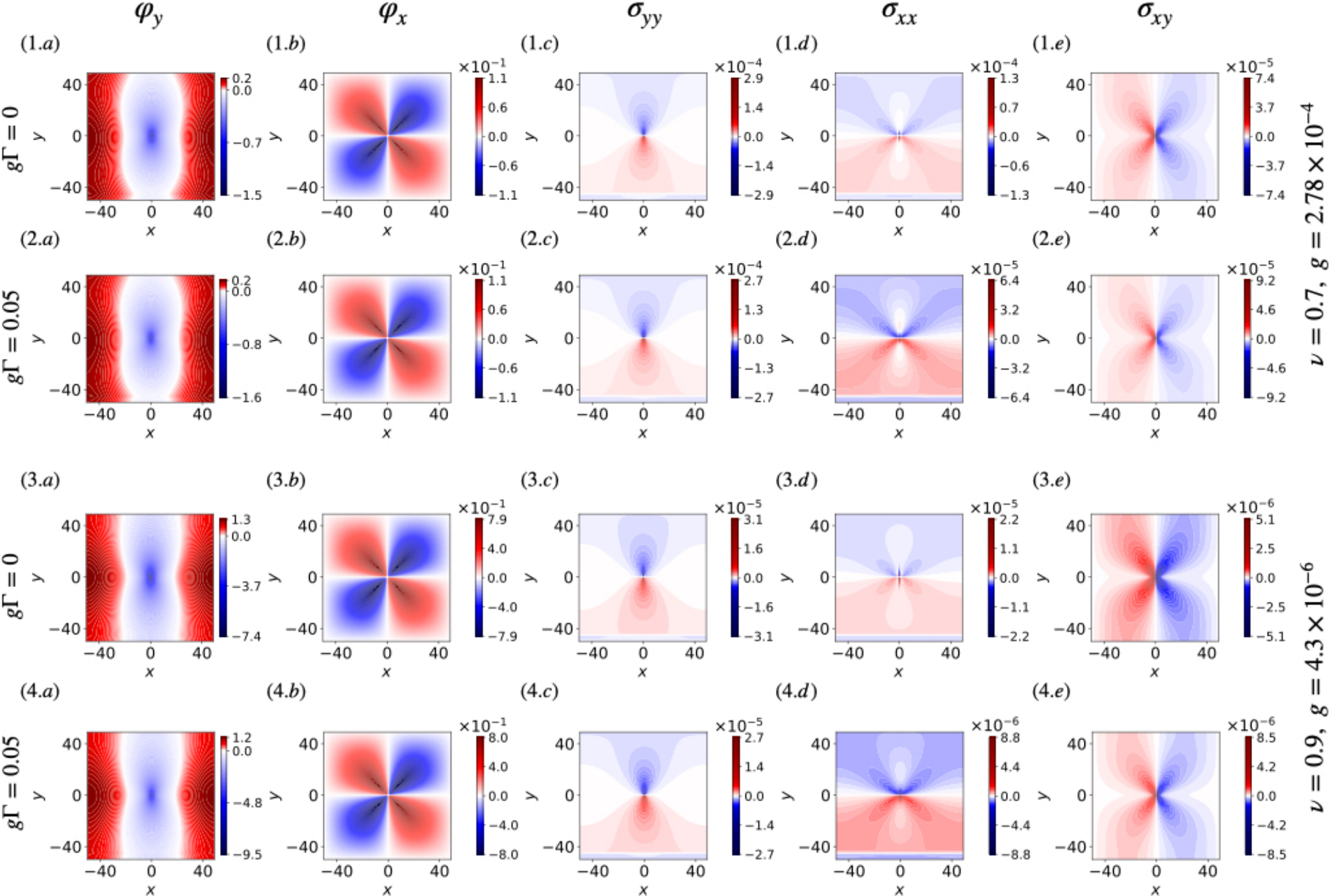}
  \caption{Theoretical response obtained by solving Eqs.~\eqref{eq_Greens_phi} and~\eqref{eq_Greens_stress} and Fourier transforming to real space. Columns (a–e) display the gauge potentials $\varphi_y$, $\varphi_x$ and the change in the stress components $\sigma_{yy}$, $\sigma_{xx}$, and $\sigma_{xy}$, respectively, for the external force geometry shown in Fig.~\ref{fig_schematic_force_geometry}. Results are presented for $p = 10^{-2}$ (top two rows) and $p = 10^{-3}$ (bottom two rows). For each pressure, the first row corresponds to the scale-independent case ($\Gamma = 0$), while the second includes a finite polarization gradient term ($g\Gamma = 0.05$). Introducing a finite $\Gamma$ suppresses stress fluctuations at small length scales and smooths sharp features in the response, thereby capturing key qualitative trends observed in simulations (Fig.~\ref{numerical_response_f_p}). The shear modulus $g$ is set from the long-wavelength plateau of the pressure–pressure correlations, and the Poisson ratio $\nu$ is fixed from the stress–response analysis presented later in this section. The gauge potentials $\varphi_i$ have been redefined to absorb the factor of $i$ in their Fourier representation.}
  \label{linear_dielectric_response}
\end{figure*}
%%%%%%%%%%%%%%%%%%%%%%%%%%%%%%%%%%%%%%%%%%%%%%%%%%%%%%%%%%%%%%%%%%%%%%%%%%%%%%%%%%%%%%%%%%%%%%%%%%%%%%%%%%%%%%%%%%%%%%%%%%%%%%%%%%%%%%%%

On the other hand, Eq.~\eqref{eq:phi_poisson} reduces to an anisotropic generalization of the screened Poisson equation for standard EM [Eq.~\eqref{eq_modified_EM_Poisson}] with $g \rightarrow \epsilon_0 $, and $\chi_{ijkl} \rightarrow \chi$. The limit $(g\chi)^{-1}\to 0$, corresponds to a high-polarizability (Debye-like) regime~[Eq.~\eqref{eq_Debye_EM}], where, in exact analogy with standard electrostatics, the gauge potential would be screened, albeit with an anisotropic screening length. The appearance of such a regime would signal a transition from dielectric screening via bound charges to screening via unbound charges. It has been proposed that such a regime exists as a precursor to unjamming~\cite{fu2025long,lemaitre2021anomalous}. 
{ The theory for this prejammed phase is based on an equation very similar to Eq.~\eqref{eq:phi_poisson} with $i\boldsymbol{\varphi}$ mapped to the displacement field (via a scalar with the dimensions of [energy][length]$^{d}$, as discussed in Section~\ref{sec_response}) emerging in response to an external force but with the $(g\chi)^{-1}$ term set to zero from the outset~\cite{lemaitre2021anomalous,procaccia2025dipole}. %

%%%%%%%%%%%%%%%%%%%%%%%%%%%%%%%%%%%%%%%%%%%%%%%%%%%%%%%%%%%%%%%%%%%%%%%%%%%%%%%%%%%%%%%%%%%%%%%%%%%%%%%%%%%%%%%%%%%%%%%%%%%%
\begin{figure*}
\centering
  \includegraphics[width=2\columnwidth]{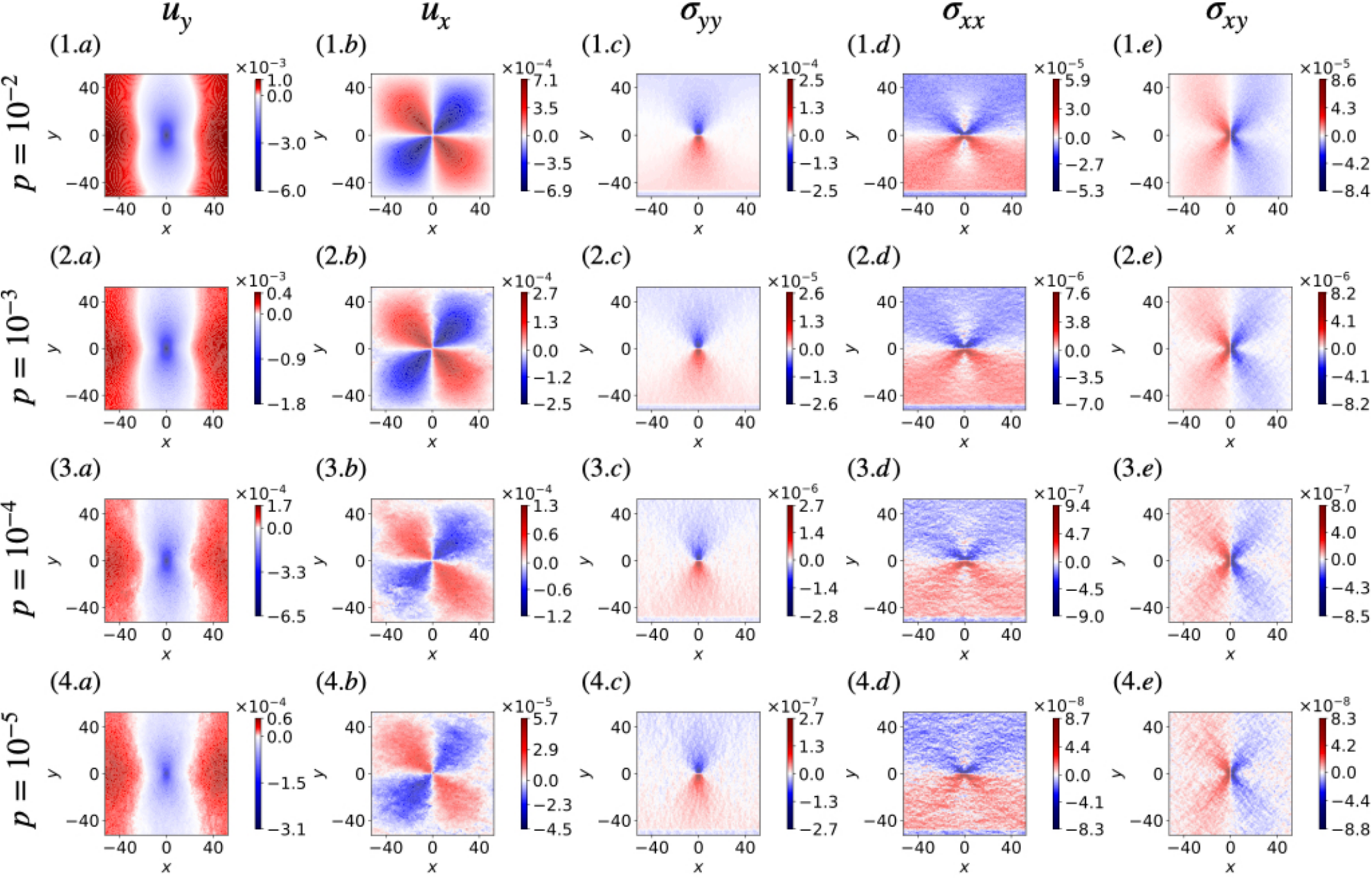}
  \caption{Disorder-averaged displacement and stress response fields from numerical simulations of jammed packings with harmonic repulsion at different pressures $p$, following the localized force geometry shown in Fig.~\ref{fig_schematic_force_geometry}. The applied body force has magnitude $10^{-1}p$ and the resulting fields are coarse-grained in unit-area boxes and averaged over $1000$ independent configurations at fixed $p$. Panels display displacement components $u_y$ and $u_x$ (a,b), and the change in the stress components $\sigma_{yy}$, $\sigma_{xx}$, and $\sigma_{xy}$ (c–e). Each row corresponds to a different pressure: (1) $p = 10^{-2}$, (2) $p = 10^{-3}$, (3) $p = 10^{-4}$, and (4) $p = 10^{-5}$. Across all pressures, the response remains smooth and featureless at large scales, consistent with scale-free elasticity, while at shorter scales it agrees closely with the predictions of the extended theory. These results demonstrate that the polarization-gradient length scale stays microscopic across all $p$ and does not generate macroscopic features near unjamming.}
  \label{numerical_response_f_p}
\end{figure*}
%%%%%%%%%%%%%%%%%%%%%%%%%%%%%%%%%%%%%%%%%%%%%%%%%%%%%%%%%%%%%%%%%%%%%%%%%%%%%%%%%%%%%%%%%%%%%%%%%%%%%%%%%%%%%%%%%%%%%%%%%%%%

As we show later in this section through the analysis of the response to external perturbations, our numerical results are inconsistent with the existence of such Debye-like screening response. Our analysis demonstrates that the Poisson ratio $\nu$ becomes weakly pressure-dependent and tends toward unity as the pressure approaches the unjamming threshold, which would imply a diverging susceptibility $\chi$ (defined in Eq.~\eqref{eq_elasticity_tensor}). However, the displacement response of jammed solids, shown in Fig.~\ref{numerical_response_f_p}, does not exhibit the corresponding screening. This can be understood by noting that $g\chi \sim g (2\nu/(1-\nu))\sim 2B\nu/(1+\nu)$ (where $B=g\frac{(1+\nu)}{(1-\nu)}$ is the Bulk modulus). It is well-known that for frictionless jamming, the bulk modulus remains finite at unjamming and decreases with pressure depending on the interaction potential between particles as the unjamming transition is approached~\cite{o2003jamming}. Thus $g\chi$ remains finite as one approaches the unjamming transition at odds with Debye screening. This feature is specific to the frictionless unjamming transition and is not generic. Indeed it would be interesting to test whether such tensorial Debye screening occurs in certain situations such as in gels~\cite{countryman2025pinch}.

%%%%%%%%%%%%%%%%%%%%%%%%%%%%%%%%%%%%%%%%%%%%%%%%%%%%%%%%%%%%%%%%%%%%%%%%%%%%%%%%%%%%%%%%%%%%%%%%%%%%%%%%%%%%%%%%%%%%%%%%%%%%%%%%%%%%%%%%
\begin{figure*}
\centering
  \includegraphics[width=2\columnwidth]{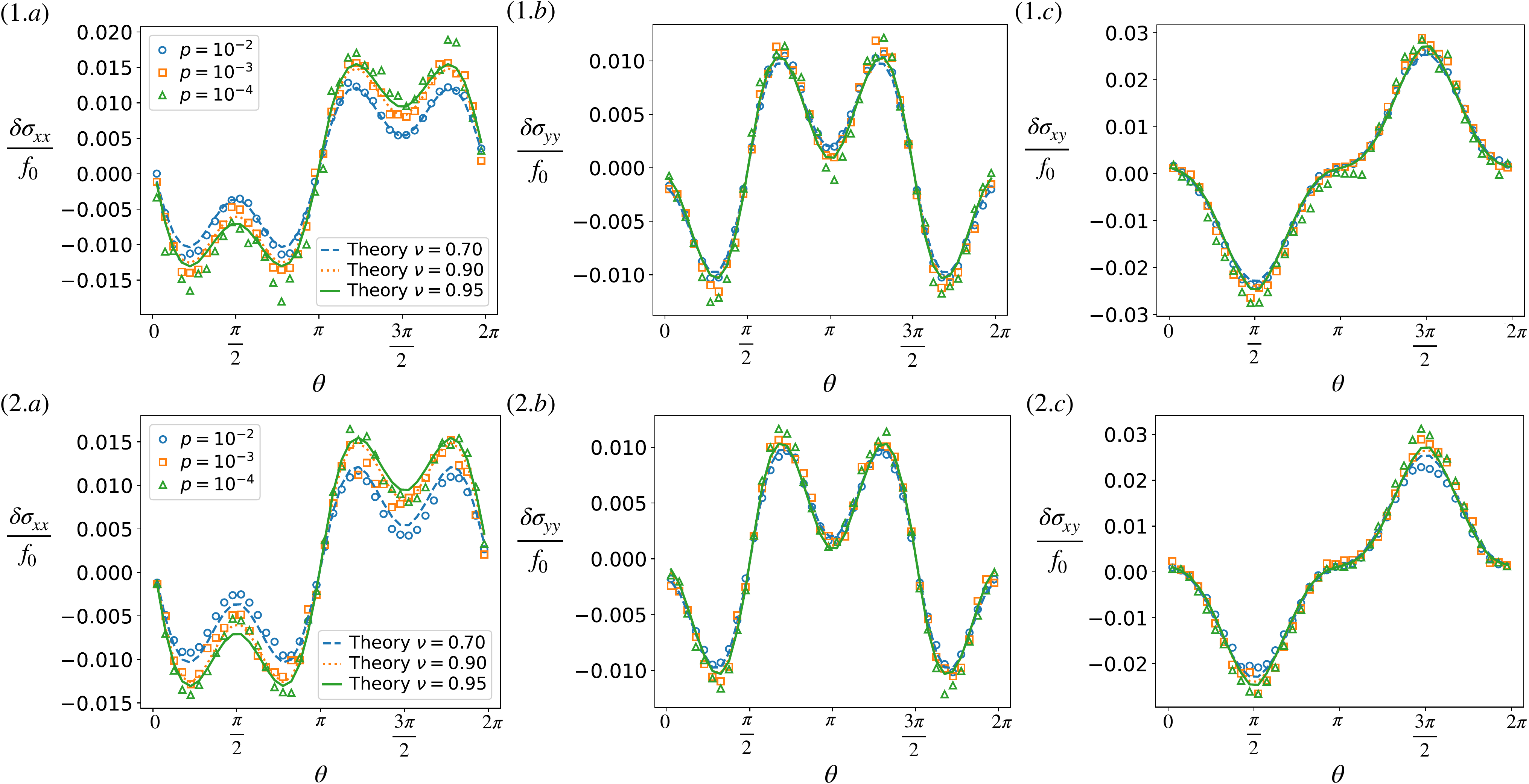}
  \caption{Angular dependence of the stress response measured in the annular region $r \in [25,35]$, shown as a function of angle $\theta$ for different pressures. Rows compare systems with harmonic (top) and Hertzian (bottom) interactions, while columns show the change in the different stress components: (a) $\sigma_{yy}$, (b) $\sigma_{xx}$, and (c) $\sigma_{xy}$. Stress responses are normalized by the applied force magnitude. Solid lines denote predictions of the theory with $\Gamma=0$, evaluated in the same annulus using the emergent Poisson ratio $\nu$ fitted at each pressure. The close agreement across pressures confirms that scale-free elasticity captures the large-scale response, with polarization-gradient effects restricted to short scales. At high pressures, the fitted $\nu$ depends on $p$, but this variation weakens at lower pressures, suggesting a limiting value of $\nu$ near unjamming. Nonlinear effects dominate very close to unjamming (see Appendix~\ref{appen_non_linear}), preventing a reliable determination of $\nu$ in this regime. The theoretical fits employ the same $\nu$ values as those obtained for the harmonic case. For Hertzian interactions, the fitted $\nu$ may differ slightly at comparable pressures.
 }
  \label{fig_stress_cuts}
\end{figure*}
%%%%%%%%%%%%%%%%%%%%%%%%%%%%%%%%%%%%%%%%%%%%%%%%%%%%%%%%%%%%%%%%%%%%%%%%%%%%%%%%%%%%%%%%%%%%%%%%%%%%%%%%%%%%%%%%%%%%%%%%%%%%%%%%%%%%%%%%

We numerically solve the response equations in Fourier space under an applied external body force, and obtain the corresponding real-space fields by inverse Fourier transform. The forcing protocol consists of a smoothed point force applied in the negative $y$-direction within a circular region of radius $2$ centered at the origin, balanced by an equal and opposite line force at $y=-45$ to ensure global force neutrality, as illustrated schematically in Fig.~\ref{fig_schematic_force_geometry}. The corresponding spatial distributions of the gauge potentials $\varphi_y$ and $\varphi_x$, together with the stress components $\sigma_{yy}$, $\sigma_{xx}$, and $\sigma_{xy}$ (columns a–e), obtained by solving Eqs.~\eqref{eq_Greens_phi} and~\eqref{eq_Greens_stress}, are presented in Fig.~\ref{linear_dielectric_response}. 

The top two rows correspond to elastic parameters representative of a system at pressure $p = 10^{-2}$, with applied force amplitude $f_0 = 10^{-3}$. The Poisson ratio $\nu$ is chosen to match the angular structure of the stress response observed in simulations (shown below), while $g$ is fixed by requiring consistency with the amplitude of the pressure–pressure correlations for the same $\nu$. The first row shows the response for $\Gamma = 0$, corresponding to the scale-free theory. The second row includes a finite polarization gradient cost with $g \Gamma = 0.05$, which introduces a characteristic screening length $\ell \sim \sqrt{g \Gamma}$ [Eq.~\eqref{eq_scale}] that smooths short-distance features in the response. The bottom two rows display results for parameters corresponding to $p = 10^{-3}$, with $f_0 = 10^{-4}$.  

While the Green’s function for the gauge potential $\varphi_i$ depends on both $g$ and $\nu$, the stress response is determined solely by $\nu$. Consequently, the effect of the microscopic length scale introduced by $\Gamma$ is most clearly visible in the stress fields. For the chosen force geometry (Fig.~\ref{fig_schematic_force_geometry}), the $\sigma_{xx}$ component is particularly sensitive to the short lengthscale regularization through $\nu$, as evident from Eq.~\eqref{eq_Greens_stress}. The inclusion of $\Gamma$ suppresses short-wavelength stress fluctuations and regularizes short-range features of the response, leading to excellent qualitative agreement with simulations. 

Fig.~\ref{numerical_response_f_p} shows the disorder-averaged mechanical response of numerically generated jammed packings with harmonic repulsion, following the perturbation protocol of Ref.~\cite{Jishnu_PRE} and extended here to low-pressure systems near the unjamming transition. While individual configurations may exhibit strong sample-to-sample fluctuations and can undergo irreversible rearrangements leading to plastic behavior (see Appendix~\ref{sec_averaging}), the disorder-averaged response remains smooth and featureless across all pressures studied, with no indication of a diverging length scale even as the system approaches the unjamming. These results also demonstrate that the large-scale mechanical behavior is well captured by the VCTG, without the need for a gradient-regularized polarization term.

We observe that the disorder-averaged displacement fields obtained from simulations and the $\varphi_i$ field are structurally similar. Importantly, the $\boldsymbol{\varphi}$ field depends on emergent elastic moduli, which are determined by internal stress fluctuations, rather than being extracted from the response of the system to boundary deformations. This suggests that, under a given set of external conditions, the VCTG gauge potential can predict the disorder-averaged relative displacement field $\bf{u}$—defined as the difference between the perturbed and unperturbed states—up to a proportionality constant. This constant depends on the external conditions and thus varies with system parameters. In the following, we demonstrate this relationship explicitly by examining how both the emergent elastic moduli and the proportionality constant relating $\varphi_i$ to $u_i$ vary with pressure in jammed solids. 

Since $\Gamma$ introduces a length scale via a scale-dependent modulus, and the numerical observations of pressure correlations and localized responses suggest that its influence is confined to short-range features. In the subsequent analysis we focus on the large-scale features of the displacement and stress fields to obtain the behavior of the emergent moduli.

We investigate the angular structure of the stress response by analyzing the radially averaged stress components within a far-field annular region, $r \in [25,35]$, as a function of the angle $\theta$ at different pressures. Fig.~\ref{fig_stress_cuts} presents these angular profiles for systems with (1) Harmonic and (2) Hertzian interactions (top and bottoms rows, respectively), showing the change in the three stress components: (a) $\sigma_{yy}$, (b) $\sigma_{xx}$, and (c) $\sigma_{xy}$. As the polarization gradient effects are negligible at these length scales and remain confined to shorter distances near the source, we compute the theoretical stress fields using $\Gamma = 0$, applying the same force geometry and evaluating the results within the same annular region. As the Green’s function for the stress response [Eq.~\eqref{eq_Greens_stress}] depends only on emergent Poisson ratio $\nu$, we extract values of $\nu$, at each pressure, which closely match the simulation results. Importantly, the VCTG describes both harmonic and Hertzian systems, underscoring its universality. 

As the system approaches unjamming, the fitted Poisson ratio exhibits progressively weaker pressure dependence, and the angular response profiles at different pressures tend to collapse. This behavior suggests that $\nu$ approaches a limiting value in the low-pressure regime. Very close to unjamming, however, nonlinear effects in the response become significant (see Appendix~\ref{appen_non_linear}), making it increasingly difficult to determine $\nu$ reliably. In this regime, the response scaled by the force magnitude does not collapse for different applied forces, indicating the need for a nonlinear generalization of the theory. Such a framework must incorporate the evolution of the emergent moduli under the nonlinear feedback from local stresses. We note that this regime is still characterized by a {\it dielectric} response at long wavelengths, and the ${\bf q}$ dependence is not affected by this nonlinearity.

%%%%%%%%%%%%%%%%%%%%%%%%%%%%%%%%%%%%%%%%%%%%%%%%%%%%%%%%%%%%%%%%%%%%
\begin{figure}
\centering
  \includegraphics[width=0.995\columnwidth]{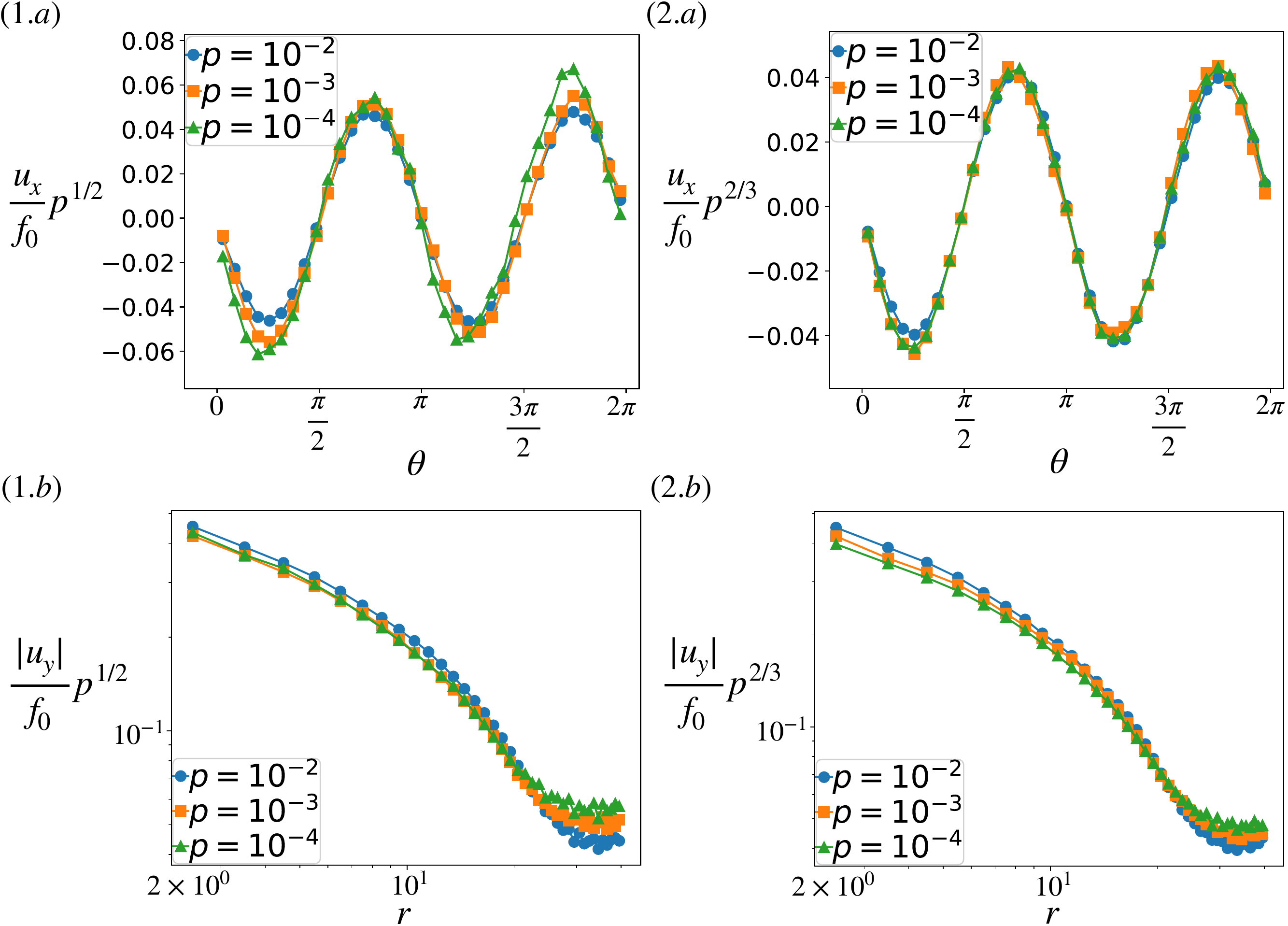}
  \caption{Displacement field components induced by the localized perturbation. Panels (1.a) and (2.a) show the angular dependence of the $x$-displacement, $u_x$, measured within the annular region $r \in [25,35]$ for systems with harmonic (1.a) and Hertzian (2.a) interactions. Displacements are plotted as a function of angle $\theta$ for different pressures, revealing anisotropy that depends on both pressure and interaction type. Panels (1.b) and (2.b) show the angularly averaged magnitude of the $y$-displacement, $|u_y|$, as a function of radial distance $r$ measured from the edge of the forcing region ($r=2$). All displacements are normalized by the magnitude of the applied point force. With this normalization, the fields exhibit pressure-dependent scaling: $u_x, u_y \sim p^{-1/2}$ for harmonic interactions and $\sim p^{-2/3}$ for Hertzian interactions. In our perturbation protocol, $u_y$ is negative near $x=0$ and positive at larger $x$, and therefore, to compare displacements at different $r$, we plot the angular average of $|u_y|$ on a logarithmic scale.
  } 
  \label{fig_displacement_cuts}
\end{figure}

As shown previously in Fig.~\ref{numerical_response_f_p}, the disorder-averaged displacement field, defined as the difference between perturbed and unperturbed configurations, shares the same spatial structure as the potential field $\boldsymbol{\varphi}$. Here we compare the scaling with pressure, of the displacement field in jammed solids against the scaling of $\boldsymbol{\varphi}$. 

As discussed earlier, the amplitude of stress correlations scales as $K_{2D}=\tfrac{g}{2}(1+\nu)\sim p^2$ near unjamming. Since $\nu$ becomes independent of pressure in this regime, it follows that $g\sim p^2$. Combined with the Green’s function for $\boldsymbol{\varphi}$ [Eq.~\eqref{eq_Greens_phi}], this yields the scaling 
\begin{equation} 
\frac{\boldsymbol{\varphi}}{\lvert \boldsymbol{f}\rvert} \sim p^{-2}. 
\end{equation}

{We now examine the displacement response obtained from simulations. In linear response, if $\mathbf{u}$ were identical to $\boldsymbol{\varphi}$, then upon normalization by the force magnitude $f_0$ it should scale as $p^{-2}$ across pressures. Fig.~\ref{fig_displacement_cuts} compares this expectation with numerical results for harmonic ($\alpha=2$) and Hertzian ($\alpha=5/2$) interactions. Panels (1.a) and (2.a) show the angular variation of $u_x(\theta)$ measured in the far field ($r \in [25,35]$), while panels (1.b) and (2.b) display the angularly averaged $|u_y(r)|$ as a function of $r$, measured from the edge of the forcing region at $r=2$, for a force applied in the $-y$ direction. The measured displacement response, once normalized by the applied force, exhibits a power-law dependence on pressure: $\mathbf{u} \sim p^{-1/2}$ for harmonic interactions and $\mathbf{u} \sim p^{-2/3}$ for Hertzian interactions. These results match the general scaling form 
\begin{equation} 
\frac{\boldsymbol{u}}{\lvert \boldsymbol{f}\rvert} \sim p^{-\frac{\alpha - 3/2}{\alpha - 1}},\label{displacement_scaling_with_p_numerics}
\end{equation} 
demonstrating that while $\mathbf{u}$ and $\boldsymbol{\varphi}$ share the same spatial structure, their pressure dependence differs. This difference can be rationalized by noting that the two fields carry distinct physical dimensions. The displacement $\mathbf{u}$ has units of length, whereas $\boldsymbol{\varphi}$, introduced as a Lagrange multiplier enforcing Gauss’s law, has units of [energy]$^{-1}$ [length]$^{-d+1}$. To relate $\boldsymbol{\varphi}$ to the physical displacement, we introduce a proportionality constant $\beta(p)$ with dimensions of [energy] [length]$^{d}$: 
\begin{equation} 
u_i = \beta(p)\varphi_i \label{eq_for_beta}.
\end{equation} 
The observed scaling of $\mathbf{u}$ then implies $\beta(p) \sim p^{-\frac{\alpha - 1/2}{\alpha - 1}}$. The factor $\beta(p)$ encodes microscopic information set by both pressure and the details of the forcing protocol. In our case, a localized force applied near the origin, the response excites shear-like, non-affine motions governed by the connectivity of the contact network and the proximity of the system to isostaticity~\cite{wyart2005rigidity}. Incorporating these isostatic scaling arguments into the energetic framework underlying Eq.~\eqref{eq_for_beta} reproduces the displacement scaling observed in simulations. Thus, once the deformation protocol and interaction potential are specified, the extended VCTG framework provides a consistent description of the displacement response in jammed solids.
}

\section{Conclusion and Discussion }
\label{sec_conclusion}

In this work, we have extended the Vector Charge Theory of Granular solids (VCTG)—a stress-only elasticity framework of prestressed solids—to incorporate both dielectric-like and Debye-type screening. Using this framework, we characterized the mechanical response of jammed packings of frictionless disks in two dimensions by analyzing stress correlations, grain displacement fields, and the scaling of the emergent elastic moduli as the system approaches the unjamming transition. 

Our main conclusions can be grouped into four categories: Foremost, using the disorder-averaged response to small external perturbations, we find no evidence of an emergent Debye-like screening length scale. If such a scale were present, it would be expected to manifest in stress correlations or mechanical response functions. Instead, the dielectric-like response persists up to the unjamming point. A distinct ``prejammed’’ phase with anomalous elasticity has been proposed in recent studies~\cite{lemaitre2021anomalous,fu2025long}, but such behavior does not emerge in our simulations. While those theoretical regimes can be formally recovered from our formulation in the Debye-like limit $(g\chi)^{-1}\to 0$, the systems we study remain in the dielectric regime all the way to unjamming. This does not rule out the possibility that other classes of prestressed solids may undergo a transition from dielectric to Debye screening. For instance, it would be interesting to explore whether such a transition occurs in gels~\cite{countryman2025pinch}.

Second, our analysis indicates that the pressure dependence of the emergent Poisson ratio becomes weaker as the system approaches the unjamming transition. This trend suggests that, in the low-pressure limit, the Poisson ratio becomes effectively pressure-independent, while the emergent shear modulus follows the scaling $g \sim p^2$. {However, the onset of non-linearities in the response near unjamming prevents a definitive determination of the pressure dependence of the elastic moduli. Capturing such effects may require extending the theory beyond the linear regime--incorporating, for instance, stress-dependent polarization or other network-level reorganizations. Within the linear regime, a dimensional analysis allows us to relate the VCTG moduli to conventional strain-based measurements, recovering the expected pressure scaling once changes in energy associated with the specific inter-particle potential and details of the deformation are taken into account. For a fixed macrostate of the ensemble, the averaged response is therefore well captured by the dielectric framework of VCTG.}

Third, we clarified the physical role of the differences of gauge potential $\boldsymbol{\varphi}$ within the VCTG framework, showing that changes in $\boldsymbol{\varphi}$-induced by external forcing under gauge fixing-encode the relative displacement field defined as the difference between the perturbed and unperturbed states. This correspondence is analogous to experimental protocols that measure ensemble-averaged responses across multiple disordered realizations under fixed macroscopic conditions. We have also demonstrated that VCTG predictions can be translated to displacement fields that arise in response to force perturbations, with applications of the analysis presented in this paper to experiments on biological tissues~\cite{charras_cell_1, charras_cell_2}, gels~\cite{goren2023probing} or frictional granular solids~\cite{mondal2022experimental}.

Finally, our results underscore the crucial role of disorder averaging. While individual realizations of amorphous packings show strong fluctuations and plastic events, ensemble averaging yields a robust, linear dielectric response that remains valid all the way up to unjamming. This statistical emergence of elasticity is reminiscent of spin-ice systems~\cite{spin_liquid}, where local constraints enforce a gauge symmetry and global behavior arises only at the ensemble level.

The theoretical framework presented in this paper is broadly applicable to disordered solids with prestress or residual stress. This gauge-theoretic, stress-only approach provides a robust formulation of prestress as arising from a screening mechanism that screens out ``charges'' that violate the divergence-free condition of the stress when external forces are present. The VCTG framework emerged out of a dielectric generalization of the vector-charge theory~\cite{pretko2017generalized}. A similar generalization of the scalar charge-theory~\cite{pretko2017generalized} would provide a useful framework for solids where prestress arises from the geometric frustration and the violation of the stress compatibility condition~\cite{Livne:2023aa,PhysRevLett.134.147401} We plan to explore this connection in the near future.

\vspace{0.8cm}
%%%%%%%%%%%%%%%%%%%

\acknowledgments

The authors acknowledge fruitful discussions with Anupam Kundu, Pinaki Chaudhuri, Madan Rao, Srikanth Sastry, Smarajit Karmakar, Itamar Procaccia, Ana{\"e}l Lema{\^\i}tre and Chandan Dasgupta. SB acknowledges funding by the Swarna Jayanti fellowship of SERB-DST (India) Grant No. SB/SJF/2021-22/12;  DST, Government of India (Nano mission), under Project No. DST/NM/TUE/QM-10/2019 (C)/7 and the Department of Atomic Energy, Government of India, under project no. RTI4001. S.C. and K.R. acknowledge support of the Department of Atomic Energy, Government of India, under Project Identification No. RTI4007. BC's work was supported by  NSF-DMR-2026834, and
NSF CBET-2228681.
%%%%%%%%%%%%%%%%%

\appendix
\section{Correlation Functions for the Modified VCTG}\label{appen_VCTG_DD_corr}

Here we provide the details the stress correlations in the VCTG augmented by the polarization gradient term [Eq.~\eqref{eq_augmented_action}]. Fourier transforming constraint action [Eq.~\eqref{MVCTG_eq_Constraint}] and integrating out the polarization field $\mathcal{P}_{ij}$ yields an effective action in terms of the stress tensor,
\begin{equation}
\begin{aligned}
\tilde{\mathcal{S}}' = \int d^d \boldsymbol{q} &\Big[
\tfrac{1}{2} \sigma_{ij}(\boldsymbol{q}) \tilde{\Lambda}_{ijkl} \sigma_{kl}(-\boldsymbol{q})
+ i \sigma_{ij}(\boldsymbol{q}) J_{ij}(-\boldsymbol{q})\\
&+ i \varphi_i(\boldsymbol{q}) \rho_i(-\boldsymbol{q}) \Big],
\end{aligned}
\end{equation}
where the emergent elasticity tensor acquires a nontrivial wavevector dependence,
\begin{equation}
\tilde{\Lambda}^{-1}_{ijkl} = g \left( \tilde{\chi}_{ijkl} + \delta_{ijkl} \right), 
\label{eq_modified_lambda_inv}
\end{equation}
where the modified susceptibility $\tilde{\chi}_{ijkl}$ is given by Eq.~\eqref{eq_modified_chi}. For brevity in notation, we suppress the explicit $\boldsymbol{q}$-dependence.

Integrating out the stress tensor $\sigma_{ij}$ yields the partition function
\begin{equation}
\mathcal{Z}[f_j^{\rm ext}] = \int [\mathcal{D}\varphi] \, e^{-S_{\rm eff}},
\end{equation}
with the effective action:
\begin{widetext}
\begin{equation}
\begin{aligned}
S_{\text{eff}}= \int d^d \boldsymbol{q}  \Big[i \varphi_i (\boldsymbol{q}) f_i^{\rm ext}(-\boldsymbol{q}) + \tfrac{1}{2}  
\big\{\mathcal{J}_{ij}(\boldsymbol{q})\tilde{\Lambda}^{-1}_{ijkl}\mathcal{J}_{kl}(-\boldsymbol{q})
+i\varphi_j(\boldsymbol{q})q_i\tilde{\Lambda}^{-1}_{ijkl}\mathcal{J}_{kl}(-\boldsymbol{q})
&-i\mathcal{J}_{ij}(\boldsymbol{q})\tilde{\Lambda}^{-1}_{ijkl}q_k\varphi_l(-\boldsymbol{q})
\\ &
+\varphi_j(\boldsymbol{q})q_i\tilde{\Lambda}^{-1}_{ijkl}q_k\varphi_l(-\boldsymbol{q})\big\} \Big].
\end{aligned}
\end{equation}

Finally, integrating out $\varphi_i$ and differentiating $\log Z$ with respect to the source fields gives the stress–stress correlation function,

\begin{equation}
\begin{aligned}
    \langle \sigma_{ij} (\boldsymbol{q}) \sigma_{kl} (-\boldsymbol{q}) \rangle_{\boldsymbol{f}^{\rm ext}=0} 
    &= \frac{\tilde{\Lambda}^{-1}_{ijkl}+\tilde{\Lambda}^{-1}_{klij}}{2} 
 - \frac{1}{4} (\tilde{\Lambda}^{-1}_{ijab}+\tilde{\Lambda}^{-1}_{abij}) 
 (\tilde{\Lambda}^{-1}_{klmn}+\tilde{\Lambda}^{-1}_{mnkl})q_bq_n \tilde{c}^{-1}_{am},
\label{eq_VCTG_Dcorr_Fourier}
\end{aligned}
\end{equation}
with $\tilde{c}_{il} = q_j  \tilde{\Lambda}^{-1}_{jikl} q_k$ introduced for compactness.
\end{widetext}
Thus, polarization gradients renormalize the stress correlations through the wavevector dependence of $\widetilde{\Lambda}^{-1}_{ijkl}$.

\vspace{0.2cm}
\noindent {\textbf{Scale-Dependent Elasticity}}

To capture the effects of spatial gradients in the polarization field, we adopt an isotropic form for the polarization stiffness tensor in Eq.~\eqref{eq_stiffness_tensor} to obtain a modified elastic tensor in Fourier space,
\begin{equation}
\begin{aligned}
    \tilde{\Lambda}^{-1}_{ijkl}(\boldsymbol{q}) &= \frac{g (2 + g \Gamma q^2)}{2(1 + g \Gamma q^2)} (\delta_{ik} \delta_{jl} + \delta_{il} \delta_{jk}) \\
    &\quad + \frac{4 g \nu}{(2 + g \Gamma q^2)(1 - \nu + (1 + 3 \nu) g \Gamma q^2)} \delta_{ij} \delta_{kl}.
\end{aligned}
\end{equation}

In two dimensions, this tensor has three eigenvalues corresponding to orthogonal modes: a compressional mode associated with the eigenvector $(1,1,0)$, and two degenerate shear modes associated with the eigenvectors $(1,-1,0)$ and $(0,0,1)$. These eigenmodes allow us to define wavevector-dependent elastic coefficients. The effective shear modulus is given in Eq.~\eqref{scale_dependent_mu}, reducing to the bare value $g$ in the long-wavelength limit ($|\boldsymbol{q}| \to 0$), and saturating to $g/2$ at short wavelengths ($|\boldsymbol{q}| \to \infty$). The compressional modulus acquires a similarly nontrivial dependence,
\begin{equation}
\tilde{\lambda}(\boldsymbol{q}) = \frac{2 g \nu}{(1 + g  \Gamma q^2) (1 - \nu + g \Gamma q^2 (1 + 3 \nu))}.
\end{equation}

At small $|\boldsymbol{q}|$, this recovers the bare value $\tilde{\lambda}(\boldsymbol{q}) \to 2g\nu/(1-\nu) \equiv \lambda$, while at large $|\boldsymbol{q}|$ it decays as $\tilde{\lambda}(\boldsymbol{q}) \sim |\boldsymbol{q}|^{-4}$. The crossover from scale-independent to scale-dependent elasticity is controlled by a characteristic length scale, $\ell = \sqrt{g \Gamma}$ [Eq.~\eqref{eq_scale}].

\vspace{0.2cm}
\noindent \textbf{Pressure Correlations and Screening} 

The explicit form of the pressure–pressure correlation function is given in Eq.~\eqref{eq_P_corr_main}. In the long-wavelength limit, this reduces to the unscreened value
\begin{equation}
    \langle P(\boldsymbol{q}) P(-\boldsymbol{q}) \rangle \;\to\; \frac{g}{2}(1 + \nu) \;\equiv\; K_{2D}.
\end{equation}
At finite $|\boldsymbol{q}|$, the correlations acquire a scale dependence, which can be expressed in terms of the effective Lam\'e coefficients as
\begin{equation}
    \langle P(\boldsymbol{q}) P(-\boldsymbol{q}) \rangle 
    = \frac{\tilde{g}(\boldsymbol{q})}{2}\big(1 + \tilde{\nu}(\boldsymbol{q})\big) 
    \;\equiv\; \widetilde{K}_{2D}(\boldsymbol{q}),
\end{equation}
with the scale-dependent Poisson ratio $\tilde{\nu}(\boldsymbol{q})$ defined in Eq.~\eqref{scale_dependent_nu}.

More generally, the full stress–stress correlation tensor exhibits anisotropic angular structure that emerges naturally within the VCTG, specifically, 
\begin{align}
    \nonumber
    &\langle \sigma_{xx}({\boldsymbol{q}}) \sigma_{xx}(-{\boldsymbol{q}}) \rangle = 4 \widetilde{K}_{2D} (\lvert{\boldsymbol{q}}\rvert) \sin^4\theta,\\
    \nonumber
    &\langle \sigma_{yy}({\boldsymbol{q}}) \sigma_{yy}(-{\boldsymbol{q}}) \rangle = 4 \widetilde{K}_{2D} (\lvert{\boldsymbol{q}}\rvert) \cos^4\theta,\\
    \nonumber
    &\langle \sigma_{xy}({\boldsymbol{q}}) \sigma_{xy}(-{\boldsymbol{q}}) \rangle = 4 \widetilde{K}_{2D} (\lvert{\boldsymbol{q}}\rvert) \sin^2\theta \cos^2\theta,\\
    \nonumber
    &\langle \sigma_{xx}({\boldsymbol{q}}) \sigma_{yy}(-{\boldsymbol{q}}) \rangle = 4 \widetilde{K}_{2D} (\lvert{\boldsymbol{q}}\rvert) \sin^2\theta \cos^2\theta,\\
    \nonumber
    &\langle \sigma_{xx}({\boldsymbol{q}}) \sigma_{xy}(-{\boldsymbol{q}}) \rangle = 4 \widetilde{K}_{2D} (\lvert{\boldsymbol{q}}\rvert) (-\sin^3\theta \cos\theta),\\
    &\langle \sigma_{xy}({\boldsymbol{q}}) \sigma_{yy}(-{\boldsymbol{q}}) \rangle = 4 \widetilde{K}_{2D} (\lvert{\boldsymbol{q}}\rvert) (-\sin\theta \cos^3\theta).
    \label{eq_stresscorr}
\end{align}

Fig.~\ref{fig_corr_fixed_mu} illustrates the Fourier-space stress–stress correlation functions, $\langle \sigma_{ij}(\boldsymbol{q}) \sigma_{kl}(-\boldsymbol{q}) \rangle$. The four columns correspond to different stress components--$C_{xxxx}$, $C_{yyyy}$, $C_{xxyy}$, and $C_{xyxy}$--and each row presents results for increasing values of the polarization stiffness $\Gamma$, while keeping the elastic parameters fixed at $\nu = 0.9$ and $g = 4 \times 10^{-6}$, consistent with the values extracted from numerical simulations at pressure $p = 10^{-3}$. For $\Gamma = 0$, corresponding to the scale-free limit of linear elasticity, the correlations are long-ranged and exhibit strong anisotropy, characteristic of unscreened theory. As $\Gamma$ is increased, short-scale screening becomes evident: correlations at large $|\boldsymbol{q}|$ are increasingly suppressed, regularizing the ultraviolet behavior of the theory. In real space, this manifests as a crossover from short-distance suppression to long-distance power-law correlations. Despite this suppression, the small-$q$ structure remains anisotropic, with pinch-point singularities that reflect directional dependence in the correlations as $|\boldsymbol{q}| \to 0$.

\section{Effects of Averaging}
\label{sec_averaging}

In disordered particle assemblies, where rigidity emerges from a complex contact network, sample-to-sample variations influence the mechanical response. These effects are particularly pronounced near the unjamming transition~\cite{o2003jamming} and near rigidity transitions~\cite{PhysRevLett.123.058001}. Fig.~\ref{fig_plastic_vs_elastic} illustrates this behavior in the displacement response of a jammed solid at two different pressures, $p$, for a system of $N=8192$ particles subjected to a point force of magnitude $0.1p$. The first two rows show the response of the same configuration to forces applied at different locations, with displacement fields, specifically the $y$-component $u_y$, plotted relative to the force application point, whereas the third row shows the average over different realizations of the perturbing force. The field is obtained by coarse-graining in square boxes of length two grain diameters, as the number of grains within each box can vary across the configuration, this introduces additional fluctuations in the coarse-grained ${\bf u}$-field.
%%%%%%%%%%%%%%%%%%%%%%%%%%%%%%%%%%%%%%%%%%%%%%%%%%%%%%%%%%%%%%%%%%%%
\begin{figure}[t!]
\centering
  \includegraphics[width=0.98\columnwidth]{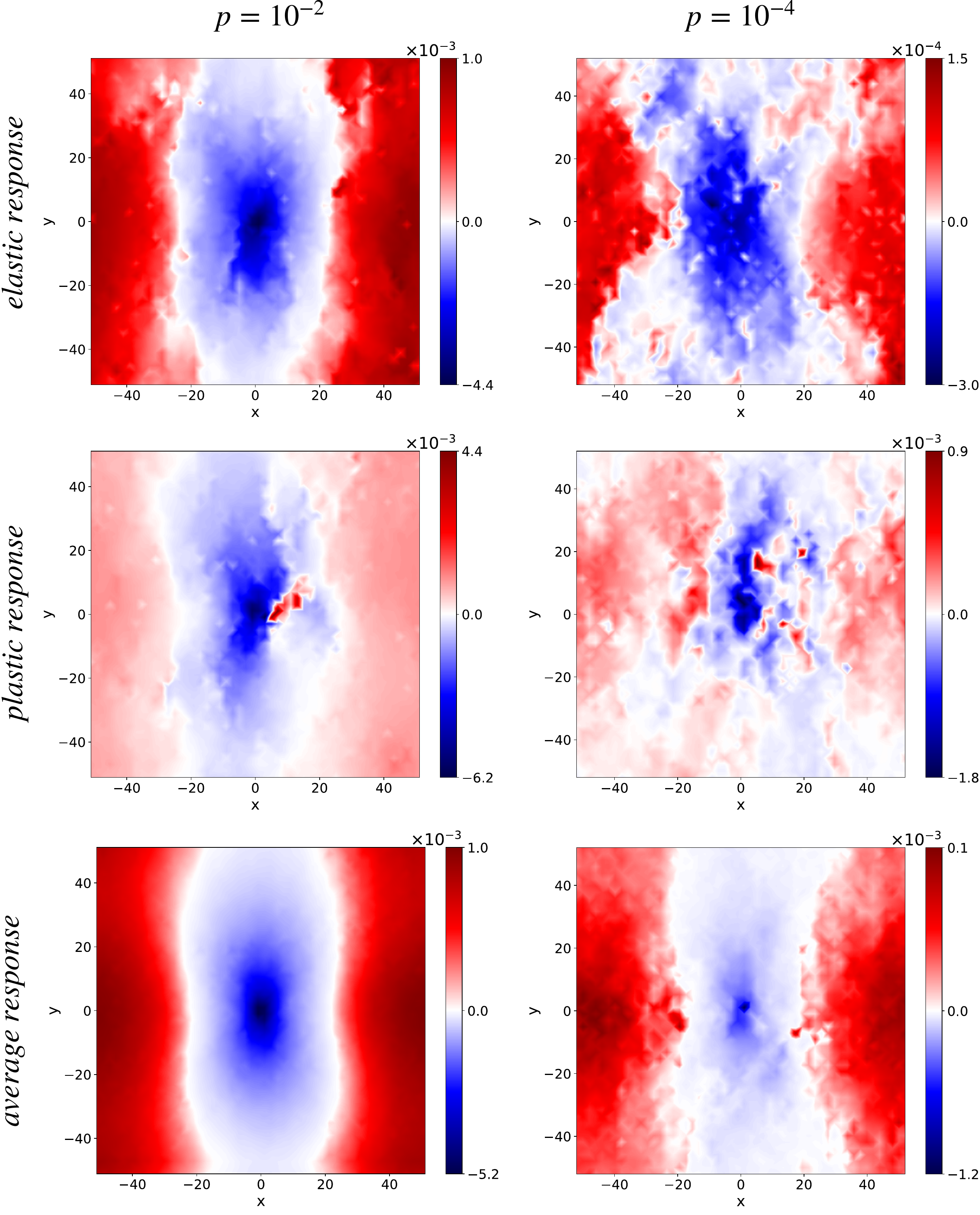}
  \caption{Displacement response of a jammed solid at pressure, $p = 10^{-2}$ ($p = 10^{-4}$) in left (right) column for a system of $N = 8192$ particles subjected to a point force of magnitude $0.1p$. The first and second rows show responses of the same configuration for two different force-application points, with displacement fields plotted relative to the force location. The third row shows displacement fields averaged over $32$ point-force realizations of the same configuration. Individual realizations may deviate strongly due to localized plastic events, but the averaged response remains elastic. At lower pressures, stronger background noise obscures the response structure, highlighting the need for averaging.}
  \label{fig_plastic_vs_elastic}
\end{figure}
%%%%%%%%%%%%%%%%%%%%%%%%%%%%%%%%%%%%%%%%%%%%%%%%%%%%%%%%%%%%%%%%%%%%

Individual realizations can either follow reversible, elastic behavior or exhibit pronounced deviations, where localized, irreversible rearrangements of the contact network produce anomalous displacement patterns. To distinguish these behaviors, we compute the mean-squared displacement (MSD), defined as $\text{MSD} = \frac{1}{N}\sum_i \lvert \mathbf{r}_f^{i}-\mathbf{r}_0^{i}\rvert^2$, where $\mathbf{r}_0^i$ denotes the position of particle $i$ in the unperturbed configuration and $\mathbf{r}_f^i$ its position after the point force is removed and the system is re-minimized. Realizations with $\text{MSD} \leq 10^{-7}$ are classified as elastic, while those with larger values correspond to plastic events. At $p=10^{-2}$, $8$ out of $32$ realizations undergo plastic instabilities, whereas at $p=10^{-4}$, $19$ out of $32$ realizations are elastic. This trend demonstrates the increasing likelihood of plastic rearrangements as the pressure is decreased. Importantly, when the displacement field is averaged over all 32 force realizations within a given configuration, the anomalous features are smoothed out, and the resulting pattern closely follows the elastic response, as shown in the third row of Fig.~\ref{fig_plastic_vs_elastic}. Near unjamming with small applied forces, the displacement field is dominated by strong background noise, and clear elastic patterns emerge only after averaging over multiple realizations, while for larger forces ($f_0 \geq p$) the response exhibits significant nonlinear effects, as discussed in Appendix~\ref{appen_non_linear}. 

While plasticity has been proposed to generate a finite screening length characteristic of a ``prejammed'' phase~\cite{lemaitre2021anomalous,fu2025long}, our results show that the ensemble-averaged responses remain well described by the dielectric screening predicted by VCTG. In our protocol—where the perturbation strength scales with pressure by applying localized forces along $-y$ to grains near the origin—plastic rearrangements occur in localized regions but are smoothed out in the disorder-averaged response, even at low pressures close to unjamming. We therefore do not observe evidence of a distinct prejammed phase with a finite, non-microscopic Debye-like screening length. The observed differences may stem from the numerical protocols used, and it remains unclear whether plastic events acquire finite weight in ensemble-averaged observables under alternative driving conditions. Our simulations clearly demonstrate that, upon averaging over multiple configurations or perturbation points, the system exhibits elastic-like behavior up to unjamming, consistent with the \textit{linear dielectric} VCTG. This behavior underscores the crucial role of disorder averaging in capturing the emergent elasticity of prestressed amorphous solids, where statistical observables become paramount. Such behavior is reminiscent of spin-ice systems~\cite{spin_liquid}, where local energetic constraints enforce Gauss's law and global properties emerge only through disorder averaging.

%%%%%%%%%%%%%%%%%%%%%%%%%%%%%%%%%%%%%%%%%%%%%%%%%%%%%%%%%%%%%%%%%%%%%%%%%%%%%%%%%%%%%%%%%%%%%%%%%%%%%%%%%%%%%%%%%%%%%%%%%%%%%%%%%%%%%%
\begin{figure}
\centering
  \includegraphics[width=0.98\columnwidth]{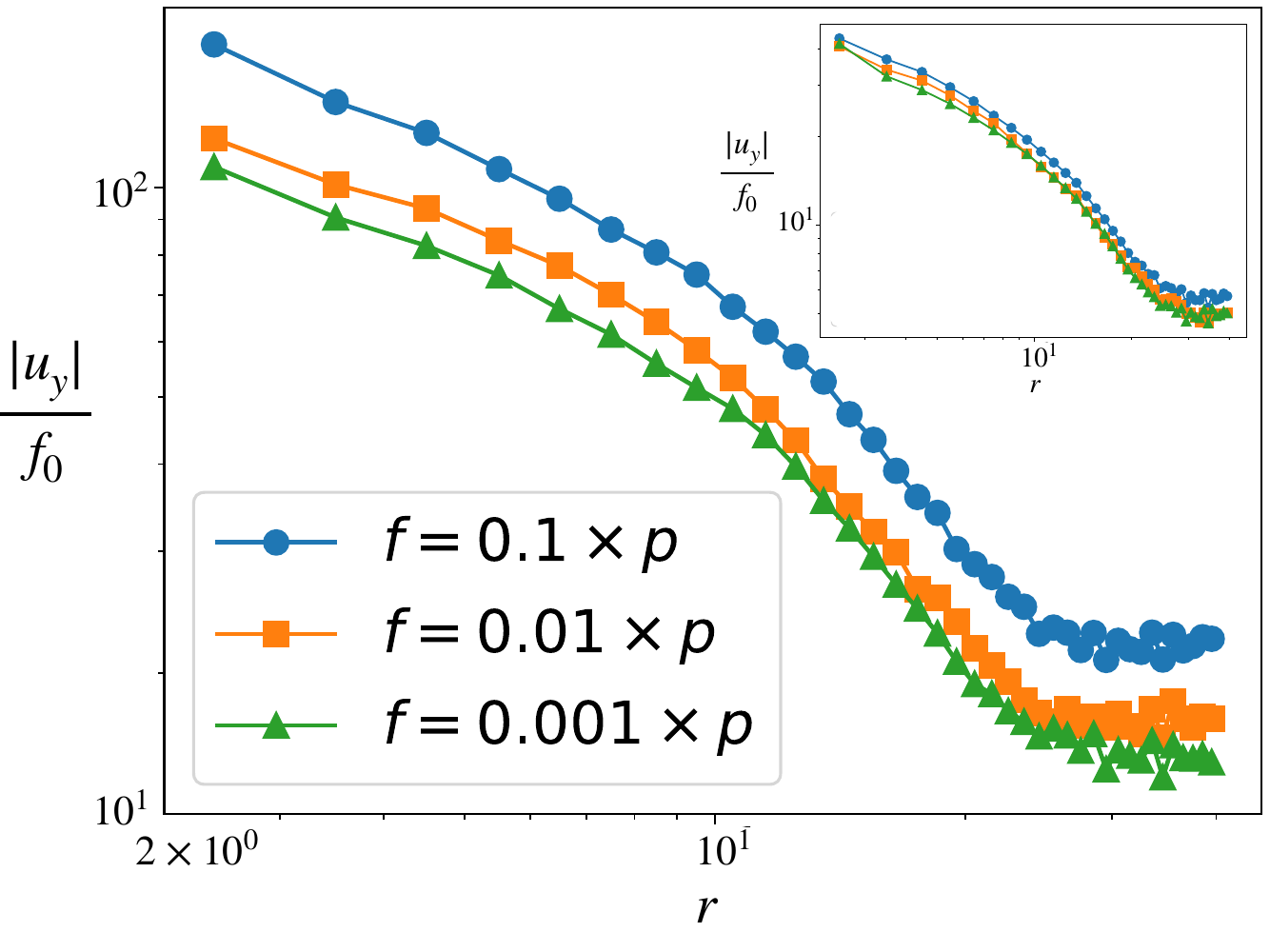}
  \caption{Deviation from linear regime near unjamming. Radial profiles of the angularly averaged absolute value of the $y$-displacement, $\lvert u_y \rvert$, normalized by the applied force magnitude $f_0$, are shown for a fixed pressure $p = 10^{-5}$ and three different forcing amplitudes: $f_0 = 0.1p$, $0.01p$, and $0.001p$. In a linear elastic medium, $u_y / f_0$ would be independent of $f_0$, resulting in collapsed curves. The observed separation between curves indicates a breakdown of linear response, suggesting that near unjamming, the system deviates from linear behavior even under infinitesimal perturbations. The inset shows the same analysis at a higher pressure $p = 10^{-4}$, where the curves nearly collapse, indicating recovery of linear behavior. In our perturbation protocol, $y$-displacement takes negative values near $x=0$ and positive values at larger $x$ even in the linear response regime described well by VCTG. To compare displacements at different $r$ on a log scale, we use the angular average of the absolute value.
  }
  \label{fig_nonlinear_cuts}
\end{figure}
%%%%%%%%%%%%%%%%%%%%%%%%%%%%%%%%%%%%%%%%%%%%%%%%%%%%%%%%%%%%%%%%%%%%%%%%%%%%%%%%%%%%%%%%%%%%%%%%%%%%%%%%%%%%%%%%%%%%%%%%%%%%%%%%%%%%%%

\begin{figure*}
\centering
  \includegraphics[width=1.9\columnwidth]{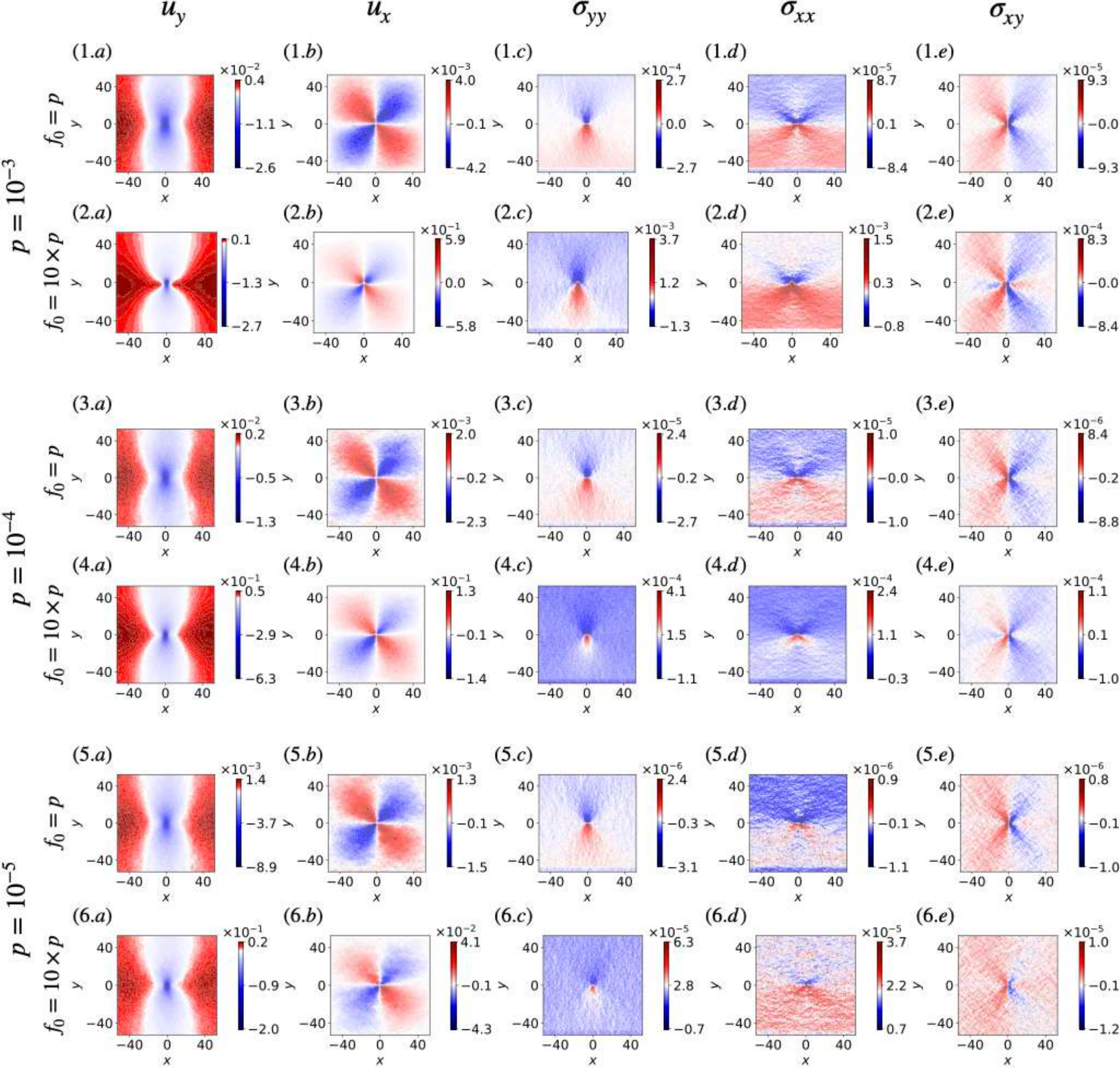}
  \caption{Response of jammed packings to a point force applied at the system center, with fields coarse-grained over boxes of side length equal to grain diameter. Each row corresponds to a different combination of pressure $p$ and applied force magnitude $f_0$: $p=10^{-3}$ with $f_0=p$ (row 1) and $f_0=10p$ (row 2); $p=10^{-4}$ with $f_0=p$ (row 3) and $f_0=10p$ (row 4); and $p=10^{-5}$ with $f_0=p$ (row 5) and $f_0=10p$ (row 6). The five columns display (a) $u_y$, (b) $u_x$, (c) $\sigma_{yy}$, (d) $\sigma_{xx}$, and (e) $\sigma_{xy}$. As shown previously (Fig.~\ref{fig_nonlinear_cuts}), the onset of nonlinear response is pressure dependent, with low-pressure systems ($p=10^{-5}$) exhibiting deviations already at $f_0 \approx 0.1p$. Here we show that this onset does not substantially modify the spatial structure: for $f_0 \leq p$ the displacement and stress fields remain nearly unchanged, while pronounced structural differences appear once the applied force exceeds the confining pressure ($f_0>p$). Importantly, this trend also persists at higher pressures, where the response maintains its form up to larger forces before showing structural modifications once $f_0>p$.
}\label{large_force_response}
\end{figure*}

%%%%%%%%%%%%%%%%%%%%%%%%%%%%%%%%%%%%%%%%%%%%%%%%%%%%%%%%%%%%%%%%%%%%%%%%% 
\section{Non-linear Effects}\label{appen_non_linear}

To probe the limits of linear regime near the unjamming transition, we systematically vary the amplitude of the applied local force and measure the resulting displacement response. Fig.~\ref{fig_nonlinear_cuts} shows the radial profiles angularly averaged $\lvert u_y(r)\rvert$, normalized by the force amplitude $f_0$, for a fixed pressure $p=10^{-5}$. In a linearly elastic medium, the normalized displacement $\lvert u_y \rvert/f_0$ should be independent of $f_0$, and all curves would collapse. Instead, we observe clear deviations from collapse even for small forcing amplitudes, indicating that the response is intrinsically nonlinear in the vicinity of unjamming. This nonlinear behavior is strongly pressure dependent: as shown in the inset for $p=10^{-4}$, the curves nearly collapse at small forces, signaling a recovery of linear response at higher pressures.

We now turn to the spatial structure of the response fields. Fig.~\ref{large_force_response} shows that the early onset of nonlinear behavior at low pressures does not substantially modify the overall patterns. For applied forces up to the confining pressure ($f_0 \leq p$), the displacement and stress fields retain their spatial structure, despite the deviations from linear scaling evident in Fig.~\ref{fig_nonlinear_cuts}. Pronounced structural changes appear only once the applied force exceeds the confining pressure ($f_0>p$). Importantly, this criterion applies across all pressures: higher-pressure systems maintain linear scaling over a broader range of forces, yet the spatial structure alters only when $f_0$ surpasses $p$.

In the following, we propose a possible theory for the non-linear effects observed in the response computation.
In the linear regime, the stress tensor $\sigma_{ij}$ relates to the unscreened electric field $E_{kl}$ via the inverse dielectric (or elastic constant) tensor $\Lambda^{-1}_{ijkl}$:  
\begin{equation}
\sigma_{ij} = \Lambda^{-1}_{ijkl} E_{kl}, \quad \text{with } E_{ij} = \frac{1}{2}(\partial_i \varphi_j + \partial_j \varphi_i),
\end{equation}
where $\varphi_i$ is the gauge potential, which we have related to the displacement field. For stress-dependent moduli, $\Lambda^{-1}_{ijkl}$ becomes a functional of the stress field:  
\begin{equation}
\Lambda^{-1}_{ijkl} \rightarrow \Lambda^{-1}_{ijkl}[\sigma],  
\end{equation}
leading to a nonlinear relation:  
\begin{equation}
\sigma_{ij} = \Lambda^{-1}_{ijkl}[\sigma] E_{kl}.  
\end{equation}
The mechanical equilibrium condition (Gauss’s law) remains:  
\begin{equation}
\partial_i \sigma_{ij} = f_j^{\text{ext}},  
\end{equation}
but the constitutive relation now implicitly depends on $\sigma_{ij}$. To close the equations, we enforce the compatibility condition (Faraday’s law in the static limit):  
\begin{equation}
\epsilon_{iak} \epsilon_{jcd} \partial_a \partial_c (\Lambda[\sigma] \sigma)_{bd} = 0,  
\end{equation}
which ensures that $E_{ij}$ derives from a gauge potential $\phi_i$.  
For small deviations from a reference stress $\sigma^{(0)}$, we expand:  
\begin{equation}
\sigma_{ij} = \sigma^{(0)}_{ij} + \delta\sigma_{ij}, \quad \Lambda^{-1}_{ijkl}[\sigma] \approx \Lambda^{(0)-1}_{ijkl} + \frac{\partial \Lambda^{-1}_{ijkl}}{\partial \sigma_{mn}} \bigg|_{\sigma^{(0)}} \delta\sigma_{mn}.  
\end{equation}
Substituting into Gauss’s law yields a Dyson-type equation:  
\begin{equation}
\partial_i \left( \Lambda^{(0)-1}_{ijkl} E_{kl} + \frac{\partial \Lambda^{-1}_{ijkl}}{\partial \sigma_{mn}} \delta\sigma_{mn} E_{kl} \right) = f_j^{\text{ext}}.  
\end{equation}
This can be solved iteratively for $\delta\sigma_{ij}$, with nonlinear corrections arising from the \(\delta\sigma\)-dependent term. A Similar construction, albeit expanding the elastic moduli in terms of the strain have been attempted in the literature~\cite{benoist2023generic}. 
We have not attempted to solve this equation, in this work, and leave its verification with numerics to future work.

\section{Standard Electrostatics with Polarization Gradient Terms}\label{Appen_EM}
In heterogeneous dielectric media, spatial inhomogeneities in structure and composition lead to fluctuating internal fields and nonuniform polarization. Variational formulations of electrostatics have addressed such situations by allowing for arbitrary spatial variations of the dielectric susceptibility~\cite{obolensky2009rigorous}. Here we explicitly extend linear dielectric theory by introducing a gradient penalty term that accounts for spatial variations in the polarization via

\begin{widetext}
    \begin{equation}
S =  \int d^d \boldsymbol{r}\Bigg[\frac{1}{2} \mathcal{P}_i \chi_{ij}^{-1} \mathcal{P}_j + \frac{1}{2} (D_i - \mathcal{P}_i)  \epsilon_{ij}^{-1} (D_j - \mathcal{P}_j) + i D_i \partial_i \psi + i \psi \rho^{\text{ext}} + \frac{\Gamma_{\alpha \beta i j}}{2} (\partial_{\alpha} \mathcal{P}_i)(\partial_{\beta} \mathcal{P}_j)\Bigg],
\label{eq_action_EM}
\end{equation}
\end{widetext}

where $\mathcal{P}_i$ is the polarization vector field, $D_i$ is the electric displacement field, and $\psi$ is a Lagrange multiplier enforcing Gauss’s law, $\partial_i D_i = \rho^{\text{ext}}$, with $\rho^{\text{ext}}$ representing the external charge density. The susceptibility and permittivity tensors are taken as $\chi_{ij}=\epsilon_0 \chi \delta_{ij}$ and $\epsilon_{ij}=\epsilon_0 \delta_{ij}$. %where $\chi$ is the electric susceptibility and $\epsilon_0$ is the vacuum permittivity. 
In the absence of the gradient term, Eq.~\eqref{eq_action_EM} reduces to the standard energy functional for a linear dielectric.

The gradient term penalizes spatial variations in the polarization field through a rank-$4$ isotropic tensor, 
\begin{equation}
\Gamma_{\alpha \beta i j}=\Gamma \left[ \delta_{\alpha \beta} \delta_{ij}+\frac{1}{2}(\delta_{\alpha i} \delta_{\beta j}+\delta_{\alpha j} \delta_{\beta i}) \right]
\end{equation}

Such an extension is particularly relevant for amorphous solids and jammed granular materials, where structural disorder and local constraints restrict polarization degrees of freedom. The gradient term introduces a characteristic length scale below which polarization fluctuations are suppressed, leading to modified electrostatic behavior. The corresponding outcome in jammed solids—where spatial variations in internal prestress are similarly penalized—is illustrated schematically in Fig.~\ref{summary_figure}.

\subsection{Correlation Functions in Fourier Space}

Transforming Eq.~\eqref{eq_action_EM} to Fourier space and integrating out the polarization field $\mathcal{P}_i$, we obtain an effective action for the displacement field $D_i$:
\begin{equation}
\begin{aligned}
\widetilde{S}' =& \int d^d \boldsymbol{q}\Bigg[\frac{1}{2} D_i(\boldsymbol{q}) \tilde{\Lambda}_{ij} D_j(-\boldsymbol{q}) + D_i(\boldsymbol{q}) q_i \psi(-\boldsymbol{q})\nonumber\\
&~~~~~~~~~~~~~~~~~~~~~~~~~~~~~~~~~~~~~~~~~~~+ i \psi(\boldsymbol{q}) \rho^{\text{ext}}(-\boldsymbol{q})\Bigg].    
\end{aligned}
\end{equation}
where the modified coupling tensor for the quadratic term is
\begin{align}
    \tilde{\Lambda}_{ij} = \epsilon_{ij}^{-1} - \epsilon_{in}^{-1} A_{nm}^{-1} \epsilon_{mj}^{-1}.
\end{align}
with $A_{ij} = \chi_{ij}^{-1} + \epsilon_{ij}^{-1} + \Gamma_{\alpha \beta i j} q_{\alpha} q_{\beta}$. To compute the correlation functions for ${D}_i$, we can now use the standard method of sources to obtain
\begin{equation}
\langle D_i D_j \rangle = \tilde{\Lambda}_{ij}^{-1} - \tilde{\Lambda}_{in}^{-1} q_n C^{-1} \tilde{\Lambda}_{jm}^{-1} q_m,
\label{eq_dd_corr_em}
\end{equation}
where $C = q_i \tilde{\Lambda}_{ij}^{-1} q_j$. For an isotropic medium the expression of $A^{-1}_{nm}$ simplifies to
\begin{equation}
    A^{-1}_{nm}=\frac{\epsilon_0 \chi}{(1+\chi)(1+\lambda^2 q^2)}\bigg[\delta_{nm}-\frac{\lambda^2}{1+2\lambda^2q^2} q_n q_m\bigg],
\end{equation}
where $\lambda^2=\frac{\epsilon_0\chi\Gamma}{1+\chi}$.
Substituting into the definition of \(\tilde{\Lambda}_{ij}\), we find the inverse modified coupling tensor:
\begin{equation}
\begin{aligned}
\tilde{\Lambda}^{-1}_{ij} =& \frac{\epsilon_0 (1 + \chi) }{1+(1+\chi)\lambda^2 q^2} \Bigg[ (1+\lambda^2 q^2)\delta_{ij}\nonumber\\  
&~~~~~~~~~~~~~~~~~~~~~~~~~~~~~~~- \frac{\chi \lambda^2}{1 + (1 + \chi)2\lambda^2 q^2 } q_i q_j \Bigg].
\end{aligned}
\end{equation}

Substituting this form into the expression for the correlation function in Eq.~\eqref{eq_dd_corr_em} yields:
\begin{equation}
\langle D_i D_j \rangle = \frac{(1+\lambda^2 q^2)}{1+(1+\chi)\lambda^2 q^2} \epsilon_0 (1 + \chi) \left[ \delta_{ij} - \frac{q_i q_j}{q^2} \right].
\label{eq:DD_corr_final}
\end{equation}
Due to the Gauss’s law constraint, $q_i D_i=0$ (for $\rho^{ext}=0)$, the displacement field correlations $\langle D_i D_j \rangle$ is strictly transverse, satisfying $q_i q_j \langle D_i D_j \rangle=0$. 
In the absence of the polarization gradient term (i.e. $\Gamma=0$), the correlation reduces to the standard form:
\begin{equation}
 \epsilon_0 (1 + \chi) \left[ \delta_{ij} - \frac{q_i q_j}{q^2} \right].
\end{equation}
 However, finite $\Gamma$ introduces a momentum-dependent amplitude through the prefactor $\frac{(1+\lambda^2 q^2)}{1+(1+\chi)\lambda^2 q^2}$. Since $\chi > 0$ this prefactor decays with increasing $q$, leading to a suppression of transverse correlations below the characteristic length scale $\lambda^2=\frac{\epsilon_0 \chi \Gamma}{1+\chi}$. 
%%%%%%%%%%%%%%%%
\subsection{Saddle point analysis}

Variation of the constrained action~\eqref{eq_action_EM} yields the following set of coupled equations of $D$, $\mathcal{P}$ and $\psi$ for the dielectric response.
\begin{align} 
& \text{(i)} \quad \partial_i D_i = \rho^{\rm ext}, \label{eq:gauss} \\ 
& \text{(ii)} \quad D_i = \mathcal{P}_i - i \epsilon_{ij} \partial_j \psi, \label{eq:D-P-relation} \\ 
& \text{(iii)} \quad \chi^{-1}_{ij} \mathcal{P}_j - \epsilon^{-1}_{ij} (D_j - \mathcal{P}_j) - \Gamma_{\alpha \beta i j} \partial_{\alpha} \partial_{\beta} \mathcal{P}_j = 0. \label{eq:P-eom}
 \end{align} 

Substituting Eq.~\eqref{eq:D-P-relation} into Eq.~\eqref{eq:P-eom} yields a modified constitutive relation for the polarization: 
\begin{equation} 
\chi^{-1}_{ij} \mathcal{P}_j - \Gamma_{\alpha \beta i j} \partial_{\alpha} \partial_{\beta} \mathcal{P}_j = -i\partial_i \psi. \label{eq:modified_P} 
\end{equation} 

In Fourier space, the polarization in an isotropic medium takes the form
\begin{equation} 
\tilde{\mathcal{P}}_i = \frac{\epsilon_0\chi}{1 +2 \epsilon_0 \chi \Gamma  q^2} q_i \tilde{\psi}.\label{eq:P_psi}
\end{equation} 

Substituting Eqs.~\eqref{eq:D-P-relation} and \eqref{eq:P_psi} into Gauss's law~\eqref{eq:gauss} leads to a modified Poisson equation: 
\begin{equation}
\left(\frac{\epsilon_0\chi q^2}{1 +2
 \epsilon_0\chi \Gamma q^2} + \epsilon_0 q^2\right) i\tilde{\psi} = \widetilde{\rho}^{ext}
\label{eq_modified_EM_Poisson}
\end{equation}

In the limit $\chi \to \infty$ corresponding to a highly polarizable medium, this equation reduces to:
\begin{equation}
\left( \frac{1}{2\Gamma} + \epsilon_0 q^2 \right) i\tilde{\psi} = \widetilde{\rho}^{\rm ext}.
\label{eq_Debye_EM}
\end{equation}

Solving Eq.~\eqref{eq_modified_EM_Poisson} yields the potential response:
\begin{equation}
i\tilde{\psi}(\boldsymbol{q}) = \frac{\widetilde{\rho}^{\rm ext}}{\epsilon_0(1 + \chi)} \left( \frac{1}{q^2} + \frac{\chi}{q^2 + (\sqrt{2}\lambda)^{-2}} \right),
\label{eq:psi_solution}
\end{equation}
where $\lambda^2 = \frac{\epsilon_0\chi \Gamma}  {(1 + \chi)}$ is the screening length. The inverse Fourier transform of the expression, in three dimensions due to a point charge is
\begin{equation}
\psi(r) = \frac{\rho^{\rm ext}}{\epsilon_0(1+\chi)} \left( \frac{1}{r} + \chi \frac{e^{-r/(\sqrt{2}\lambda)}}{r} \right),
\end{equation}
where the second term reflects Yukawa-type screening with decay length $\lambda$. In the limit $\chi\gg 1$, this reduces to
\begin{equation}
\psi(r) = \frac{\rho^{\rm ext}}{\epsilon_0} \frac{e^{-r/(\sqrt{2}\lambda)}}{r}, \quad \text{with} \quad \lambda = \sqrt{\epsilon_0 \Gamma}.
\end{equation}
while in two dimensions, the corresponding expression is
\begin{equation}
\psi(r) = \frac{\rho^{\rm ext}}{2\pi \epsilon_0(1+\chi)} \left[ -\log(r) +\chi K_0\left(\frac{r}{\sqrt{2}\lambda}\right) \right],
\end{equation}
where $K_0$ is the modified Bessel function of the second kind. This function exhibits a crossover from logarithmic behavior at short distances to exponential decay at large distances.

\bibliographystyle{apsrev4-2}
\bibliography{Response_Bibliography}

\begin{thebibliography}{59}%
\makeatletter
\providecommand \@ifxundefined [1]{%
 \@ifx{#1\undefined}
}%
\providecommand \@ifnum [1]{%
 \ifnum #1\expandafter \@firstoftwo
 \else \expandafter \@secondoftwo
 \fi
}%
\providecommand \@ifx [1]{%
 \ifx #1\expandafter \@firstoftwo
 \else \expandafter \@secondoftwo
 \fi
}%
\providecommand \natexlab [1]{#1}%
\providecommand \enquote  [1]{``#1''}%
\providecommand \bibnamefont  [1]{#1}%
\providecommand \bibfnamefont [1]{#1}%
\providecommand \citenamefont [1]{#1}%
\providecommand \href@noop [0]{\@secondoftwo}%
\providecommand \href [0]{\begingroup \@sanitize@url \@href}%
\providecommand \@href[1]{\@@startlink{#1}\@@href}%
\providecommand \@@href[1]{\endgroup#1\@@endlink}%
\providecommand \@sanitize@url [0]{\catcode `\\12\catcode `\$12\catcode `\&12\catcode `\#12\catcode `\^12\catcode `\_12\catcode `\%12\relax}%
\providecommand \@@startlink[1]{}%
\providecommand \@@endlink[0]{}%
\providecommand \url  [0]{\begingroup\@sanitize@url \@url }%
\providecommand \@url [1]{\endgroup\@href {#1}{\urlprefix }}%
\providecommand \urlprefix  [0]{URL }%
\providecommand \Eprint [0]{\href }%
\providecommand \doibase [0]{https://doi.org/}%
\providecommand \selectlanguage [0]{\@gobble}%
\providecommand \bibinfo  [0]{\@secondoftwo}%
\providecommand \bibfield  [0]{\@secondoftwo}%
\providecommand \translation [1]{[#1]}%
\providecommand \BibitemOpen [0]{}%
\providecommand \bibitemStop [0]{}%
\providecommand \bibitemNoStop [0]{.\EOS\space}%
\providecommand \EOS [0]{\spacefactor3000\relax}%
\providecommand \BibitemShut  [1]{\csname bibitem#1\endcsname}%
\let\auto@bib@innerbib\@empty
%</preamble>
\bibitem [{\citenamefont {Landau}\ \emph {et~al.}(2012)\citenamefont {Landau}, \citenamefont {Pitaevskii}, \citenamefont {Kosevich},\ and\ \citenamefont {Lifshitz}}]{landau2012theory}%
  \BibitemOpen
  \bibfield  {author} {\bibinfo {author} {\bibfnamefont {L.~D.}\ \bibnamefont {Landau}}, \bibinfo {author} {\bibfnamefont {L.}~\bibnamefont {Pitaevskii}}, \bibinfo {author} {\bibfnamefont {A.~M.}\ \bibnamefont {Kosevich}},\ and\ \bibinfo {author} {\bibfnamefont {E.~M.}\ \bibnamefont {Lifshitz}},\ }\href@noop {} {\emph {\bibinfo {title} {Theory of elasticity: volume 7}}},\ Vol.~\bibinfo {volume} {7}\ (\bibinfo  {publisher} {Elsevier},\ \bibinfo {year} {2012})\BibitemShut {NoStop}%
\bibitem [{\citenamefont {Alexander}(1998)}]{alexander1998amorphous}%
  \BibitemOpen
  \bibfield  {author} {\bibinfo {author} {\bibfnamefont {S.}~\bibnamefont {Alexander}},\ }\href {https://doi.org/https://doi.org/10.1016/S0370-1573(97)00069-0} {\bibfield  {journal} {\bibinfo  {journal} {Physics reports}\ }\textbf {\bibinfo {volume} {296}},\ \bibinfo {pages} {65} (\bibinfo {year} {1998})}\BibitemShut {NoStop}%
\bibitem [{\citenamefont {Cates}\ \emph {et~al.}(1998)\citenamefont {Cates}, \citenamefont {Wittmer}, \citenamefont {Bouchaud},\ and\ \citenamefont {Claudin}}]{cates1998jamming}%
  \BibitemOpen
  \bibfield  {author} {\bibinfo {author} {\bibfnamefont {M.}~\bibnamefont {Cates}}, \bibinfo {author} {\bibfnamefont {J.}~\bibnamefont {Wittmer}}, \bibinfo {author} {\bibfnamefont {J.-P.}\ \bibnamefont {Bouchaud}},\ and\ \bibinfo {author} {\bibfnamefont {P.}~\bibnamefont {Claudin}},\ }\href {https://doi.org/https://doi.org/10.1103/PhysRevLett.81.1841} {\bibfield  {journal} {\bibinfo  {journal} {Physical review letters}\ }\textbf {\bibinfo {volume} {81}},\ \bibinfo {pages} {1841} (\bibinfo {year} {1998})}\BibitemShut {NoStop}%
\bibitem [{\citenamefont {Bouchaud}(2004)}]{bouchaud2004course}%
  \BibitemOpen
  \bibfield  {author} {\bibinfo {author} {\bibfnamefont {J.}~\bibnamefont {Bouchaud}},\ }in\ \href {https://arxiv.org/pdf/cond-mat/0211196} {\emph {\bibinfo {booktitle} {Slow Relaxations and nonequilibrium dynamics in condensed matter: Les Houches Session LXXVII, 1-26 July, 2002}}}\ (\bibinfo  {publisher} {Springer},\ \bibinfo {year} {2004})\ pp.\ \bibinfo {pages} {131--197}\BibitemShut {NoStop}%
\bibitem [{\citenamefont {Bi}\ \emph {et~al.}(2011)\citenamefont {Bi}, \citenamefont {Zhang}, \citenamefont {Chakraborty},\ and\ \citenamefont {Behringer}}]{bi2011jamming}%
  \BibitemOpen
  \bibfield  {author} {\bibinfo {author} {\bibfnamefont {D.}~\bibnamefont {Bi}}, \bibinfo {author} {\bibfnamefont {J.}~\bibnamefont {Zhang}}, \bibinfo {author} {\bibfnamefont {B.}~\bibnamefont {Chakraborty}},\ and\ \bibinfo {author} {\bibfnamefont {R.~P.}\ \bibnamefont {Behringer}},\ }\href {https://doi.org/https://doi.org/10.1038/nature10667} {\bibfield  {journal} {\bibinfo  {journal} {Nature}\ }\textbf {\bibinfo {volume} {480}},\ \bibinfo {pages} {355} (\bibinfo {year} {2011})}\BibitemShut {NoStop}%
\bibitem [{\citenamefont {Brown}\ and\ \citenamefont {Jaeger}(2014)}]{brown2014shear}%
  \BibitemOpen
  \bibfield  {author} {\bibinfo {author} {\bibfnamefont {E.}~\bibnamefont {Brown}}\ and\ \bibinfo {author} {\bibfnamefont {H.~M.}\ \bibnamefont {Jaeger}},\ }\href {https://doi.org/10.1088/0034-4885/77/4/046602} {\bibfield  {journal} {\bibinfo  {journal} {Reports on Progress in Physics}\ }\textbf {\bibinfo {volume} {77}},\ \bibinfo {pages} {046602} (\bibinfo {year} {2014})}\BibitemShut {NoStop}%
\bibitem [{\citenamefont {Lema{\^\i}tre}(2017)}]{lemaitre2017inherent}%
  \BibitemOpen
  \bibfield  {author} {\bibinfo {author} {\bibfnamefont {A.}~\bibnamefont {Lema{\^\i}tre}},\ }\href {https://doi.org/https://doi.org/10.1103/PhysRevE.96.052101} {\bibfield  {journal} {\bibinfo  {journal} {Physical Review E}\ }\textbf {\bibinfo {volume} {96}},\ \bibinfo {pages} {052101} (\bibinfo {year} {2017})}\BibitemShut {NoStop}%
\bibitem [{\citenamefont {Tong}\ \emph {et~al.}(2020)\citenamefont {Tong}, \citenamefont {Sengupta},\ and\ \citenamefont {Tanaka}}]{tong2020emergent}%
  \BibitemOpen
  \bibfield  {author} {\bibinfo {author} {\bibfnamefont {H.}~\bibnamefont {Tong}}, \bibinfo {author} {\bibfnamefont {S.}~\bibnamefont {Sengupta}},\ and\ \bibinfo {author} {\bibfnamefont {H.}~\bibnamefont {Tanaka}},\ }\href {https://doi.org/https://doi.org/10.1038/s41467-020-18663-7} {\bibfield  {journal} {\bibinfo  {journal} {Nature communications}\ }\textbf {\bibinfo {volume} {11}},\ \bibinfo {pages} {4863} (\bibinfo {year} {2020})}\BibitemShut {NoStop}%
\bibitem [{\citenamefont {Zhang}\ \emph {et~al.}(2022)\citenamefont {Zhang}, \citenamefont {Stanifer}, \citenamefont {Vasisht}, \citenamefont {Zhang}, \citenamefont {Del~Gado},\ and\ \citenamefont {Mao}}]{zhang2022prestressed}%
  \BibitemOpen
  \bibfield  {author} {\bibinfo {author} {\bibfnamefont {S.}~\bibnamefont {Zhang}}, \bibinfo {author} {\bibfnamefont {E.}~\bibnamefont {Stanifer}}, \bibinfo {author} {\bibfnamefont {V.~V.}\ \bibnamefont {Vasisht}}, \bibinfo {author} {\bibfnamefont {L.}~\bibnamefont {Zhang}}, \bibinfo {author} {\bibfnamefont {E.}~\bibnamefont {Del~Gado}},\ and\ \bibinfo {author} {\bibfnamefont {X.}~\bibnamefont {Mao}},\ }\href {https://doi.org/10.1103/PhysRevResearch.4.043181} {\bibfield  {journal} {\bibinfo  {journal} {Physical Review Research}\ }\textbf {\bibinfo {volume} {4}},\ \bibinfo {pages} {043181} (\bibinfo {year} {2022})}\BibitemShut {NoStop}%
\bibitem [{\citenamefont {Pica~Ciamarra}\ and\ \citenamefont {Coniglio}(2009)}]{pica2009jamming}%
  \BibitemOpen
  \bibfield  {author} {\bibinfo {author} {\bibfnamefont {M.}~\bibnamefont {Pica~Ciamarra}}\ and\ \bibinfo {author} {\bibfnamefont {A.}~\bibnamefont {Coniglio}},\ }\href {https://doi.org/https://doi.org/10.1103/PhysRevLett.103.235701} {\bibfield  {journal} {\bibinfo  {journal} {Physical review letters}\ }\textbf {\bibinfo {volume} {103}},\ \bibinfo {pages} {235701} (\bibinfo {year} {2009})}\BibitemShut {NoStop}%
\bibitem [{\citenamefont {Mao}\ \emph {et~al.}(2009)\citenamefont {Mao}, \citenamefont {Goldbart}, \citenamefont {Xing},\ and\ \citenamefont {Zippelius}}]{Mao2009}%
  \BibitemOpen
  \bibfield  {author} {\bibinfo {author} {\bibfnamefont {X.}~\bibnamefont {Mao}}, \bibinfo {author} {\bibfnamefont {P.~M.}\ \bibnamefont {Goldbart}}, \bibinfo {author} {\bibfnamefont {X.}~\bibnamefont {Xing}},\ and\ \bibinfo {author} {\bibfnamefont {A.}~\bibnamefont {Zippelius}},\ }\href {https://doi.org/10.1103/PhysRevE.80.031140} {\bibfield  {journal} {\bibinfo  {journal} {Physical Review E}\ }\textbf {\bibinfo {volume} {80}},\ \bibinfo {pages} {031140} (\bibinfo {year} {2009})}\BibitemShut {NoStop}%
\bibitem [{\citenamefont {Bouzid}\ \emph {et~al.}(2017)\citenamefont {Bouzid}, \citenamefont {Colombo}, \citenamefont {Barbosa},\ and\ \citenamefont {Del~Gado}}]{bouzid_elastically_2017}%
  \BibitemOpen
  \bibfield  {author} {\bibinfo {author} {\bibfnamefont {M.}~\bibnamefont {Bouzid}}, \bibinfo {author} {\bibfnamefont {J.}~\bibnamefont {Colombo}}, \bibinfo {author} {\bibfnamefont {L.~V.}\ \bibnamefont {Barbosa}},\ and\ \bibinfo {author} {\bibfnamefont {E.}~\bibnamefont {Del~Gado}},\ }\href {https://doi.org/10.1038/ncomms15846} {\bibfield  {journal} {\bibinfo  {journal} {Nature Communications}\ }\textbf {\bibinfo {volume} {8}},\ \bibinfo {pages} {15846} (\bibinfo {year} {2017})},\ \bibinfo {note} {publisher: Springer Nature}\BibitemShut {NoStop}%
\bibitem [{\citenamefont {Del~Gado}(2021)}]{delgado2021}%
  \BibitemOpen
  \bibfield  {author} {\bibinfo {author} {\bibfnamefont {E.}~\bibnamefont {Del~Gado}},\ }in\ \href {https://doi.org/10.1007/978-3-642-27737-5_734-1} {\emph {\bibinfo {booktitle} {Encyclopedia of Complexity and Systems Science}}}\ (\bibinfo  {publisher} {Springer International Publishing},\ \bibinfo {address} {Cham},\ \bibinfo {year} {2021})\BibitemShut {NoStop}%
\bibitem [{\citenamefont {Geng}\ \emph {et~al.}(2001)\citenamefont {Geng}, \citenamefont {Howell}, \citenamefont {Longhi}, \citenamefont {Behringer}, \citenamefont {Reydellet}, \citenamefont {Vanel}, \citenamefont {Cl{\'e}ment},\ and\ \citenamefont {Luding}}]{geng2001footprints}%
  \BibitemOpen
  \bibfield  {author} {\bibinfo {author} {\bibfnamefont {J.}~\bibnamefont {Geng}}, \bibinfo {author} {\bibfnamefont {D.}~\bibnamefont {Howell}}, \bibinfo {author} {\bibfnamefont {E.}~\bibnamefont {Longhi}}, \bibinfo {author} {\bibfnamefont {R.}~\bibnamefont {Behringer}}, \bibinfo {author} {\bibfnamefont {G.}~\bibnamefont {Reydellet}}, \bibinfo {author} {\bibfnamefont {L.}~\bibnamefont {Vanel}}, \bibinfo {author} {\bibfnamefont {E.}~\bibnamefont {Cl{\'e}ment}},\ and\ \bibinfo {author} {\bibfnamefont {S.}~\bibnamefont {Luding}},\ }\href {https://doi.org/https://doi.org/10.1103/PhysRevLett.87.035506} {\bibfield  {journal} {\bibinfo  {journal} {Physical Review Letters}\ }\textbf {\bibinfo {volume} {87}},\ \bibinfo {pages} {035506} (\bibinfo {year} {2001})}\BibitemShut {NoStop}%
\bibitem [{\citenamefont {Otto}\ \emph {et~al.}(2003)\citenamefont {Otto}, \citenamefont {Bouchaud}, \citenamefont {Claudin},\ and\ \citenamefont {Socolar}}]{otto2003anisotropy}%
  \BibitemOpen
  \bibfield  {author} {\bibinfo {author} {\bibfnamefont {M.}~\bibnamefont {Otto}}, \bibinfo {author} {\bibfnamefont {J.-P.}\ \bibnamefont {Bouchaud}}, \bibinfo {author} {\bibfnamefont {P.}~\bibnamefont {Claudin}},\ and\ \bibinfo {author} {\bibfnamefont {J.~E.}\ \bibnamefont {Socolar}},\ }\href {https://doi.org/https://doi.org/10.1103/PhysRevE.67.031302} {\bibfield  {journal} {\bibinfo  {journal} {Physical Review E}\ }\textbf {\bibinfo {volume} {67}},\ \bibinfo {pages} {031302} (\bibinfo {year} {2003})}\BibitemShut {NoStop}%
\bibitem [{\citenamefont {Schuh}\ and\ \citenamefont {Lund}(2003)}]{schuh2003atomistic}%
  \BibitemOpen
  \bibfield  {author} {\bibinfo {author} {\bibfnamefont {C.~A.}\ \bibnamefont {Schuh}}\ and\ \bibinfo {author} {\bibfnamefont {A.~C.}\ \bibnamefont {Lund}},\ }\href {https://doi.org/https://doi.org/10.1038/nmat918} {\bibfield  {journal} {\bibinfo  {journal} {Nature materials}\ }\textbf {\bibinfo {volume} {2}},\ \bibinfo {pages} {449} (\bibinfo {year} {2003})}\BibitemShut {NoStop}%
\bibitem [{\citenamefont {Maloney}\ and\ \citenamefont {Lemaitre}(2006)}]{maloney2006amorphous}%
  \BibitemOpen
  \bibfield  {author} {\bibinfo {author} {\bibfnamefont {C.~E.}\ \bibnamefont {Maloney}}\ and\ \bibinfo {author} {\bibfnamefont {A.}~\bibnamefont {Lemaitre}},\ }\href {https://doi.org/https://doi.org/10.1103/PhysRevE.74.016118} {\bibfield  {journal} {\bibinfo  {journal} {Physical Review E}\ }\textbf {\bibinfo {volume} {74}},\ \bibinfo {pages} {016118} (\bibinfo {year} {2006})}\BibitemShut {NoStop}%
\bibitem [{\citenamefont {Liu}\ and\ \citenamefont {Nagel}(1998)}]{liu1998jamming}%
  \BibitemOpen
  \bibfield  {author} {\bibinfo {author} {\bibfnamefont {A.~J.}\ \bibnamefont {Liu}}\ and\ \bibinfo {author} {\bibfnamefont {S.~R.}\ \bibnamefont {Nagel}},\ }\href {https://doi.org/https://doi.org/10.1038/23819} {\bibfield  {journal} {\bibinfo  {journal} {Nature}\ }\textbf {\bibinfo {volume} {396}},\ \bibinfo {pages} {21} (\bibinfo {year} {1998})}\BibitemShut {NoStop}%
\bibitem [{\citenamefont {O'Hern}\ \emph {et~al.}(2002)\citenamefont {O'Hern}, \citenamefont {Langer}, \citenamefont {Liu},\ and\ \citenamefont {Nagel}}]{o2002random}%
  \BibitemOpen
  \bibfield  {author} {\bibinfo {author} {\bibfnamefont {C.~S.}\ \bibnamefont {O'Hern}}, \bibinfo {author} {\bibfnamefont {S.~A.}\ \bibnamefont {Langer}}, \bibinfo {author} {\bibfnamefont {A.~J.}\ \bibnamefont {Liu}},\ and\ \bibinfo {author} {\bibfnamefont {S.~R.}\ \bibnamefont {Nagel}},\ }\href {https://doi.org/10.1103/PhysRevLett.88.075507} {\bibfield  {journal} {\bibinfo  {journal} {Physical Review Letters}\ }\textbf {\bibinfo {volume} {88}},\ \bibinfo {pages} {075507} (\bibinfo {year} {2002})}\BibitemShut {NoStop}%
\bibitem [{\citenamefont {O'hern}\ \emph {et~al.}(2003)\citenamefont {O'hern}, \citenamefont {Silbert}, \citenamefont {Liu},\ and\ \citenamefont {Nagel}}]{o2003jamming}%
  \BibitemOpen
  \bibfield  {author} {\bibinfo {author} {\bibfnamefont {C.~S.}\ \bibnamefont {O'hern}}, \bibinfo {author} {\bibfnamefont {L.~E.}\ \bibnamefont {Silbert}}, \bibinfo {author} {\bibfnamefont {A.~J.}\ \bibnamefont {Liu}},\ and\ \bibinfo {author} {\bibfnamefont {S.~R.}\ \bibnamefont {Nagel}},\ }\href {https://doi.org/10.1103/PhysRevE.68.011306} {\bibfield  {journal} {\bibinfo  {journal} {Physical Review E}\ }\textbf {\bibinfo {volume} {68}},\ \bibinfo {pages} {011306} (\bibinfo {year} {2003})}\BibitemShut {NoStop}%
\bibitem [{\citenamefont {Trappe}\ \emph {et~al.}(2001)\citenamefont {Trappe}, \citenamefont {Prasad}, \citenamefont {Cipelletti}, \citenamefont {Segre},\ and\ \citenamefont {Weitz}}]{trappe2001jamming}%
  \BibitemOpen
  \bibfield  {author} {\bibinfo {author} {\bibfnamefont {V.}~\bibnamefont {Trappe}}, \bibinfo {author} {\bibfnamefont {V.}~\bibnamefont {Prasad}}, \bibinfo {author} {\bibfnamefont {L.}~\bibnamefont {Cipelletti}}, \bibinfo {author} {\bibfnamefont {P.}~\bibnamefont {Segre}},\ and\ \bibinfo {author} {\bibfnamefont {D.~A.}\ \bibnamefont {Weitz}},\ }\href {https://doi.org/https://doi.org/10.1038/35081021} {\bibfield  {journal} {\bibinfo  {journal} {Nature}\ }\textbf {\bibinfo {volume} {411}},\ \bibinfo {pages} {772} (\bibinfo {year} {2001})}\BibitemShut {NoStop}%
\bibitem [{\citenamefont {Gopal}\ and\ \citenamefont {Durian}(2003)}]{gopal2003relaxing}%
  \BibitemOpen
  \bibfield  {author} {\bibinfo {author} {\bibfnamefont {A.}~\bibnamefont {Gopal}}\ and\ \bibinfo {author} {\bibfnamefont {D.~J.}\ \bibnamefont {Durian}},\ }\href {https://doi.org/10.1103/PhysRevLett.91.188303} {\bibfield  {journal} {\bibinfo  {journal} {Physical review letters}\ }\textbf {\bibinfo {volume} {91}},\ \bibinfo {pages} {188303} (\bibinfo {year} {2003})}\BibitemShut {NoStop}%
\bibitem [{\citenamefont {Chaudhuri}\ \emph {et~al.}(2007)\citenamefont {Chaudhuri}, \citenamefont {Berthier},\ and\ \citenamefont {Kob}}]{chaudhuri2007universal}%
  \BibitemOpen
  \bibfield  {author} {\bibinfo {author} {\bibfnamefont {P.}~\bibnamefont {Chaudhuri}}, \bibinfo {author} {\bibfnamefont {L.}~\bibnamefont {Berthier}},\ and\ \bibinfo {author} {\bibfnamefont {W.}~\bibnamefont {Kob}},\ }\href {https://doi.org/10.1103/PhysRevLett.99.060604} {\bibfield  {journal} {\bibinfo  {journal} {Physical review letters}\ }\textbf {\bibinfo {volume} {99}},\ \bibinfo {pages} {060604} (\bibinfo {year} {2007})}\BibitemShut {NoStop}%
\bibitem [{\citenamefont {Nampoothiri}\ \emph {et~al.}(2020)\citenamefont {Nampoothiri}, \citenamefont {Wang}, \citenamefont {Ramola}, \citenamefont {Zhang}, \citenamefont {Bhattacharjee},\ and\ \citenamefont {Chakraborty}}]{Jishnu_PRL}%
  \BibitemOpen
  \bibfield  {author} {\bibinfo {author} {\bibfnamefont {J.~N.}\ \bibnamefont {Nampoothiri}}, \bibinfo {author} {\bibfnamefont {Y.}~\bibnamefont {Wang}}, \bibinfo {author} {\bibfnamefont {K.}~\bibnamefont {Ramola}}, \bibinfo {author} {\bibfnamefont {J.}~\bibnamefont {Zhang}}, \bibinfo {author} {\bibfnamefont {S.}~\bibnamefont {Bhattacharjee}},\ and\ \bibinfo {author} {\bibfnamefont {B.}~\bibnamefont {Chakraborty}},\ }\href {https://doi.org/10.1103/PhysRevLett.125.118002} {\bibfield  {journal} {\bibinfo  {journal} {Physical review letters}\ }\textbf {\bibinfo {volume} {125}},\ \bibinfo {pages} {118002} (\bibinfo {year} {2020})}\BibitemShut {NoStop}%
\bibitem [{\citenamefont {Nampoothiri}\ \emph {et~al.}(2022)\citenamefont {Nampoothiri}, \citenamefont {D'Eon}, \citenamefont {Ramola}, \citenamefont {Chakraborty},\ and\ \citenamefont {Bhattacharjee}}]{Jishnu_PRE}%
  \BibitemOpen
  \bibfield  {author} {\bibinfo {author} {\bibfnamefont {J.~N.}\ \bibnamefont {Nampoothiri}}, \bibinfo {author} {\bibfnamefont {M.}~\bibnamefont {D'Eon}}, \bibinfo {author} {\bibfnamefont {K.}~\bibnamefont {Ramola}}, \bibinfo {author} {\bibfnamefont {B.}~\bibnamefont {Chakraborty}},\ and\ \bibinfo {author} {\bibfnamefont {S.}~\bibnamefont {Bhattacharjee}},\ }\href {https://doi.org/10.1103/PhysRevE.106.065004} {\bibfield  {journal} {\bibinfo  {journal} {Physical Review E}\ }\textbf {\bibinfo {volume} {106}},\ \bibinfo {pages} {065004} (\bibinfo {year} {2022})},\ \bibinfo {note} {see Supplemental Material at \url{https://journals.aps.org/pre/supplemental/10.1103/PhysRevE.106.065004}}\BibitemShut {NoStop}%
\bibitem [{\citenamefont {Pretko}(2017)}]{pretko2017generalized}%
  \BibitemOpen
  \bibfield  {author} {\bibinfo {author} {\bibfnamefont {M.}~\bibnamefont {Pretko}},\ }\href {https://doi.org/10.1103/PhysRevB.96.035119} {\bibfield  {journal} {\bibinfo  {journal} {Physical Review B}\ }\textbf {\bibinfo {volume} {96}},\ \bibinfo {pages} {035119} (\bibinfo {year} {2017})}\BibitemShut {NoStop}%
\bibitem [{\citenamefont {Jackson}(2021)}]{jackson2021classical}%
  \BibitemOpen
  \bibfield  {author} {\bibinfo {author} {\bibfnamefont {J.~D.}\ \bibnamefont {Jackson}},\ }\href@noop {} {\emph {\bibinfo {title} {Classical electrodynamics}}}\ (\bibinfo  {publisher} {John Wiley \& Sons},\ \bibinfo {year} {2021})\BibitemShut {NoStop}%
\bibitem [{\citenamefont {Lema{\^\i}tre}\ \emph {et~al.}(2021)\citenamefont {Lema{\^\i}tre}, \citenamefont {Mondal}, \citenamefont {Moshe}, \citenamefont {Procaccia}, \citenamefont {Roy},\ and\ \citenamefont {Screiber-Re'em}}]{lemaitre2021anomalous}%
  \BibitemOpen
  \bibfield  {author} {\bibinfo {author} {\bibfnamefont {A.}~\bibnamefont {Lema{\^\i}tre}}, \bibinfo {author} {\bibfnamefont {C.}~\bibnamefont {Mondal}}, \bibinfo {author} {\bibfnamefont {M.}~\bibnamefont {Moshe}}, \bibinfo {author} {\bibfnamefont {I.}~\bibnamefont {Procaccia}}, \bibinfo {author} {\bibfnamefont {S.}~\bibnamefont {Roy}},\ and\ \bibinfo {author} {\bibfnamefont {K.}~\bibnamefont {Screiber-Re'em}},\ }\href {https://doi.org/https://doi.org/10.1103/PhysRevE.104.024904} {\bibfield  {journal} {\bibinfo  {journal} {Physical Review E}\ }\textbf {\bibinfo {volume} {104}},\ \bibinfo {pages} {024904} (\bibinfo {year} {2021})}\BibitemShut {NoStop}%
\bibitem [{\citenamefont {Kumar}\ \emph {et~al.}(2022)\citenamefont {Kumar}, \citenamefont {Moshe}, \citenamefont {Procaccia},\ and\ \citenamefont {Singh}}]{Kumar:2022aa}%
  \BibitemOpen
  \bibfield  {author} {\bibinfo {author} {\bibfnamefont {A.}~\bibnamefont {Kumar}}, \bibinfo {author} {\bibfnamefont {M.}~\bibnamefont {Moshe}}, \bibinfo {author} {\bibfnamefont {I.}~\bibnamefont {Procaccia}},\ and\ \bibinfo {author} {\bibfnamefont {M.}~\bibnamefont {Singh}},\ }\href {https://doi.org/10.1103/PhysRevE.106.015001} {\bibfield  {journal} {\bibinfo  {journal} {Physical Review E}\ }\textbf {\bibinfo {volume} {106}},\ \bibinfo {pages} {015001} (\bibinfo {year} {2022})}\BibitemShut {NoStop}%
\bibitem [{\citenamefont {Fu}\ \emph {et~al.}(2025)\citenamefont {Fu}, \citenamefont {Jin}, \citenamefont {Pan},\ and\ \citenamefont {Procaccia}}]{fu2025long}%
  \BibitemOpen
  \bibfield  {author} {\bibinfo {author} {\bibfnamefont {Y.}~\bibnamefont {Fu}}, \bibinfo {author} {\bibfnamefont {Y.}~\bibnamefont {Jin}}, \bibinfo {author} {\bibfnamefont {D.}~\bibnamefont {Pan}},\ and\ \bibinfo {author} {\bibfnamefont {I.}~\bibnamefont {Procaccia}},\ }\href {https://doi.org/10.1103/PhysRevLett.134.178201} {\bibfield  {journal} {\bibinfo  {journal} {Physical Review Letters}\ }\textbf {\bibinfo {volume} {134}},\ \bibinfo {pages} {178201} (\bibinfo {year} {2025})}\BibitemShut {NoStop}%
\bibitem [{\citenamefont {Lois}\ \emph {et~al.}(2009)\citenamefont {Lois}, \citenamefont {Zhang}, \citenamefont {Majmudar}, \citenamefont {Henkes}, \citenamefont {Chakraborty}, \citenamefont {O'Hern},\ and\ \citenamefont {Behringer}}]{lois2009stress}%
  \BibitemOpen
  \bibfield  {author} {\bibinfo {author} {\bibfnamefont {G.}~\bibnamefont {Lois}}, \bibinfo {author} {\bibfnamefont {J.}~\bibnamefont {Zhang}}, \bibinfo {author} {\bibfnamefont {T.}~\bibnamefont {Majmudar}}, \bibinfo {author} {\bibfnamefont {S.}~\bibnamefont {Henkes}}, \bibinfo {author} {\bibfnamefont {B.}~\bibnamefont {Chakraborty}}, \bibinfo {author} {\bibfnamefont {C.}~\bibnamefont {O'Hern}},\ and\ \bibinfo {author} {\bibfnamefont {R.}~\bibnamefont {Behringer}},\ }\href {https://doi.org/http://dx.doi.org/10.1103/PhysRevE.80.060303} {\bibfield  {journal} {\bibinfo  {journal} {Physical Review E}\ }\textbf {\bibinfo {volume} {80}},\ \bibinfo {pages} {060303} (\bibinfo {year} {2009})}\BibitemShut {NoStop}%
\bibitem [{\citenamefont {Lema{\^\i}tre}(2018)}]{lemaitre2018stress}%
  \BibitemOpen
  \bibfield  {author} {\bibinfo {author} {\bibfnamefont {A.}~\bibnamefont {Lema{\^\i}tre}},\ }\href {https://doi.org/https://doi.org/10.1063/1.5041461} {\bibfield  {journal} {\bibinfo  {journal} {The Journal of Chemical Physics}\ }\textbf {\bibinfo {volume} {149}},\ \bibinfo {pages} {104107} (\bibinfo {year} {2018})}\BibitemShut {NoStop}%
\bibitem [{\citenamefont {DeGiuli}(2018)}]{degiuli2018field}%
  \BibitemOpen
  \bibfield  {author} {\bibinfo {author} {\bibfnamefont {E.}~\bibnamefont {DeGiuli}},\ }\href {https://doi.org/10.1103/PhysRevLett.121.118001} {\bibfield  {journal} {\bibinfo  {journal} {Physical review letters}\ }\textbf {\bibinfo {volume} {121}},\ \bibinfo {pages} {118001} (\bibinfo {year} {2018})}\BibitemShut {NoStop}%
\bibitem [{\citenamefont {Henkes}\ and\ \citenamefont {Chakraborty}(2009)}]{henkes2009statistical}%
  \BibitemOpen
  \bibfield  {author} {\bibinfo {author} {\bibfnamefont {S.}~\bibnamefont {Henkes}}\ and\ \bibinfo {author} {\bibfnamefont {B.}~\bibnamefont {Chakraborty}},\ }\href {https://doi.org/10.1103/PhysRevE.79.061301} {\bibfield  {journal} {\bibinfo  {journal} {Physical Review E}\ }\textbf {\bibinfo {volume} {79}},\ \bibinfo {pages} {061301} (\bibinfo {year} {2009})}\BibitemShut {NoStop}%
\bibitem [{\citenamefont {McNamara}\ \emph {et~al.}(2016)\citenamefont {McNamara}, \citenamefont {Crassous},\ and\ \citenamefont {Amon}}]{mcnamara2016eshelby}%
  \BibitemOpen
  \bibfield  {author} {\bibinfo {author} {\bibfnamefont {S.}~\bibnamefont {McNamara}}, \bibinfo {author} {\bibfnamefont {J.}~\bibnamefont {Crassous}},\ and\ \bibinfo {author} {\bibfnamefont {A.}~\bibnamefont {Amon}},\ }\href {https://doi.org/10.1103/PhysRevE.94.022907} {\bibfield  {journal} {\bibinfo  {journal} {Physical Review E}\ }\textbf {\bibinfo {volume} {94}},\ \bibinfo {pages} {022907} (\bibinfo {year} {2016})}\BibitemShut {NoStop}%
\bibitem [{\citenamefont {Wang}\ \emph {et~al.}(2020)\citenamefont {Wang}, \citenamefont {Wang},\ and\ \citenamefont {Zhang}}]{wang2020connecting}%
  \BibitemOpen
  \bibfield  {author} {\bibinfo {author} {\bibfnamefont {Y.}~\bibnamefont {Wang}}, \bibinfo {author} {\bibfnamefont {Y.}~\bibnamefont {Wang}},\ and\ \bibinfo {author} {\bibfnamefont {J.}~\bibnamefont {Zhang}},\ }\href {https://doi.org/https://doi.org/10.1038/s41467-020-18217-x} {\bibfield  {journal} {\bibinfo  {journal} {Nature communications}\ }\textbf {\bibinfo {volume} {11}},\ \bibinfo {pages} {4349} (\bibinfo {year} {2020})}\BibitemShut {NoStop}%
\bibitem [{\citenamefont {Vinutha}\ \emph {et~al.}(2023)\citenamefont {Vinutha}, \citenamefont {Diaz~Ruiz}, \citenamefont {Mao}, \citenamefont {Chakraborty},\ and\ \citenamefont {Del~Gado}}]{vinutha2023stress}%
  \BibitemOpen
  \bibfield  {author} {\bibinfo {author} {\bibfnamefont {H.}~\bibnamefont {Vinutha}}, \bibinfo {author} {\bibfnamefont {F.~D.}\ \bibnamefont {Diaz~Ruiz}}, \bibinfo {author} {\bibfnamefont {X.}~\bibnamefont {Mao}}, \bibinfo {author} {\bibfnamefont {B.}~\bibnamefont {Chakraborty}},\ and\ \bibinfo {author} {\bibfnamefont {E.}~\bibnamefont {Del~Gado}},\ }\href {https://doi.org/https://doi.org/10.1063/5.0131473} {\bibfield  {journal} {\bibinfo  {journal} {The Journal of chemical physics}\ }\textbf {\bibinfo {volume} {158}},\ \bibinfo {pages} {114104} (\bibinfo {year} {2023})}\BibitemShut {NoStop}%
\bibitem [{\citenamefont {Countryman}\ \emph {et~al.}(2025)\citenamefont {Countryman}, \citenamefont {Vinutha}, \citenamefont {Ruiz}, \citenamefont {Mao}, \citenamefont {Del~Gado},\ and\ \citenamefont {Chakraborty}}]{countryman2025pinch}%
  \BibitemOpen
  \bibfield  {author} {\bibinfo {author} {\bibfnamefont {A.}~\bibnamefont {Countryman}}, \bibinfo {author} {\bibfnamefont {H.}~\bibnamefont {Vinutha}}, \bibinfo {author} {\bibfnamefont {F.~D.}\ \bibnamefont {Ruiz}}, \bibinfo {author} {\bibfnamefont {X.}~\bibnamefont {Mao}}, \bibinfo {author} {\bibfnamefont {E.}~\bibnamefont {Del~Gado}},\ and\ \bibinfo {author} {\bibfnamefont {B.}~\bibnamefont {Chakraborty}},\ }\href {https://doi.org/10.1039/d5sm00296f} {\bibfield  {journal} {\bibinfo  {journal} {Soft matter}\ }\textbf {\bibinfo {volume} {21}},\ \bibinfo {pages} {4812} (\bibinfo {year} {2025})}\BibitemShut {NoStop}%
\bibitem [{\citenamefont {Maharana}\ \emph {et~al.}(2024)\citenamefont {Maharana}, \citenamefont {Das}, \citenamefont {Chaudhuri},\ and\ \citenamefont {Ramola}}]{maharana2024universal}%
  \BibitemOpen
  \bibfield  {author} {\bibinfo {author} {\bibfnamefont {R.}~\bibnamefont {Maharana}}, \bibinfo {author} {\bibfnamefont {D.}~\bibnamefont {Das}}, \bibinfo {author} {\bibfnamefont {P.}~\bibnamefont {Chaudhuri}},\ and\ \bibinfo {author} {\bibfnamefont {K.}~\bibnamefont {Ramola}},\ }\href {https://doi.org/10.1103/PhysRevE.109.044903} {\bibfield  {journal} {\bibinfo  {journal} {Physical Review E}\ }\textbf {\bibinfo {volume} {109}},\ \bibinfo {pages} {044903} (\bibinfo {year} {2024})}\BibitemShut {NoStop}%
\bibitem [{\citenamefont {Maharana}\ and\ \citenamefont {Ramola}(2024)}]{maharana2024stress}%
  \BibitemOpen
  \bibfield  {author} {\bibinfo {author} {\bibfnamefont {R.}~\bibnamefont {Maharana}}\ and\ \bibinfo {author} {\bibfnamefont {K.}~\bibnamefont {Ramola}},\ }\href {https://doi.org/10.21468/SciPostPhys.17.1.012} {\bibfield  {journal} {\bibinfo  {journal} {SciPost Physics}\ }\textbf {\bibinfo {volume} {17}},\ \bibinfo {pages} {012} (\bibinfo {year} {2024})}\BibitemShut {NoStop}%
\bibitem [{\citenamefont {Ball}\ and\ \citenamefont {Blumenfeld}(2002)}]{ball2002stress}%
  \BibitemOpen
  \bibfield  {author} {\bibinfo {author} {\bibfnamefont {R.~C.}\ \bibnamefont {Ball}}\ and\ \bibinfo {author} {\bibfnamefont {R.}~\bibnamefont {Blumenfeld}},\ }\href {https://doi.org/https://doi.org/10.1103/PhysRevLett.88.115505} {\bibfield  {journal} {\bibinfo  {journal} {Physical review letters}\ }\textbf {\bibinfo {volume} {88}},\ \bibinfo {pages} {115505} (\bibinfo {year} {2002})}\BibitemShut {NoStop}%
\bibitem [{\citenamefont {Henkes}\ and\ \citenamefont {Chakraborty}(2005)}]{henkes2005jamming}%
  \BibitemOpen
  \bibfield  {author} {\bibinfo {author} {\bibfnamefont {S.}~\bibnamefont {Henkes}}\ and\ \bibinfo {author} {\bibfnamefont {B.}~\bibnamefont {Chakraborty}},\ }\href {https://doi.org/https://doi.org/10.1103/PhysRevLett.95.198002} {\bibfield  {journal} {\bibinfo  {journal} {Physical review letters}\ }\textbf {\bibinfo {volume} {95}},\ \bibinfo {pages} {198002} (\bibinfo {year} {2005})}\BibitemShut {NoStop}%
\bibitem [{\citenamefont {Ramola}\ and\ \citenamefont {Chakraborty}(2017)}]{ramola2017stress}%
  \BibitemOpen
  \bibfield  {author} {\bibinfo {author} {\bibfnamefont {K.}~\bibnamefont {Ramola}}\ and\ \bibinfo {author} {\bibfnamefont {B.}~\bibnamefont {Chakraborty}},\ }\href {https://doi.org/https://doi.org/10.1007/s10955-017-1857-0} {\bibfield  {journal} {\bibinfo  {journal} {Journal of Statistical Physics}\ }\textbf {\bibinfo {volume} {169}},\ \bibinfo {pages} {1} (\bibinfo {year} {2017})}\BibitemShut {NoStop}%
\bibitem [{\citenamefont {Majmudar}\ and\ \citenamefont {Behringer}(2005)}]{majmudar2005contact}%
  \BibitemOpen
  \bibfield  {author} {\bibinfo {author} {\bibfnamefont {T.~S.}\ \bibnamefont {Majmudar}}\ and\ \bibinfo {author} {\bibfnamefont {R.~P.}\ \bibnamefont {Behringer}},\ }\href {https://doi.org/https://doi.org/10.1038/nature03805} {\bibfield  {journal} {\bibinfo  {journal} {nature}\ }\textbf {\bibinfo {volume} {435}},\ \bibinfo {pages} {1079} (\bibinfo {year} {2005})}\BibitemShut {NoStop}%
\bibitem [{\citenamefont {Behringer}\ and\ \citenamefont {Chakraborty}(2019)}]{Behringer:2019aa}%
  \BibitemOpen
  \bibfield  {author} {\bibinfo {author} {\bibfnamefont {R.~P.}\ \bibnamefont {Behringer}}\ and\ \bibinfo {author} {\bibfnamefont {B.}~\bibnamefont {Chakraborty}},\ }\href {https://doi.org/10.1088/1361-6633/aadc3c} {\bibfield  {journal} {\bibinfo  {journal} {Rep Prog Phys}\ }\textbf {\bibinfo {volume} {82}},\ \bibinfo {pages} {012601} (\bibinfo {year} {2019})}\BibitemShut {NoStop}%
\bibitem [{\citenamefont {Mondal}\ \emph {et~al.}(2022)\citenamefont {Mondal}, \citenamefont {Moshe}, \citenamefont {Procaccia}, \citenamefont {Roy}, \citenamefont {Shang},\ and\ \citenamefont {Zhang}}]{mondal2022experimental}%
  \BibitemOpen
  \bibfield  {author} {\bibinfo {author} {\bibfnamefont {C.}~\bibnamefont {Mondal}}, \bibinfo {author} {\bibfnamefont {M.}~\bibnamefont {Moshe}}, \bibinfo {author} {\bibfnamefont {I.}~\bibnamefont {Procaccia}}, \bibinfo {author} {\bibfnamefont {S.}~\bibnamefont {Roy}}, \bibinfo {author} {\bibfnamefont {J.}~\bibnamefont {Shang}},\ and\ \bibinfo {author} {\bibfnamefont {J.}~\bibnamefont {Zhang}},\ }\href {https://doi.org/https://doi.org/10.1016/j.chaos.2022.112609} {\bibfield  {journal} {\bibinfo  {journal} {Chaos, Solitons \& Fractals}\ }\textbf {\bibinfo {volume} {164}},\ \bibinfo {pages} {112609} (\bibinfo {year} {2022})}\BibitemShut {NoStop}%
\bibitem [{\citenamefont {Chakraborty}\ and\ \citenamefont {Ramola}(2024)}]{chakraborty2024long}%
  \BibitemOpen
  \bibfield  {author} {\bibinfo {author} {\bibfnamefont {S.}~\bibnamefont {Chakraborty}}\ and\ \bibinfo {author} {\bibfnamefont {K.}~\bibnamefont {Ramola}},\ }\href {https://doi.org/10.1039/d4sm00328d} {\bibfield  {journal} {\bibinfo  {journal} {Soft Matter}\ }\textbf {\bibinfo {volume} {20}},\ \bibinfo {pages} {4895} (\bibinfo {year} {2024})}\BibitemShut {NoStop}%
\bibitem [{\citenamefont {Ciarlet}(1994)}]{ciarlet1994three}%
  \BibitemOpen
  \bibfield  {author} {\bibinfo {author} {\bibfnamefont {P.~G.}\ \bibnamefont {Ciarlet}},\ }\href@noop {} {\emph {\bibinfo {title} {Three-dimensional elasticity}}},\ Vol.~\bibinfo {volume} {20}\ (\bibinfo  {publisher} {Elsevier},\ \bibinfo {year} {1994})\BibitemShut {NoStop}%
\bibitem [{\citenamefont {Procaccia}\ and\ \citenamefont {Samanta}(2025)}]{procaccia2025dipole}%
  \BibitemOpen
  \bibfield  {author} {\bibinfo {author} {\bibfnamefont {I.}~\bibnamefont {Procaccia}}\ and\ \bibinfo {author} {\bibfnamefont {T.}~\bibnamefont {Samanta}},\ }\href {https://doi.org/https://doi.org/10.1073/pnas.2427273122} {\bibfield  {journal} {\bibinfo  {journal} {Proceedings of the National Academy of Sciences}\ }\textbf {\bibinfo {volume} {122}},\ \bibinfo {pages} {e2427273122} (\bibinfo {year} {2025})}\BibitemShut {NoStop}%
\bibitem [{\citenamefont {Wyart}(2005)}]{wyart2005rigidity}%
  \BibitemOpen
  \bibfield  {author} {\bibinfo {author} {\bibfnamefont {M.}~\bibnamefont {Wyart}},\ }\bibfield  {journal} {\bibinfo  {journal} {arXiv preprint cond-mat/0512155}\ }\href {https://doi.org/https://arxiv.org/pdf/cond-mat/0512155.pdf} {https://arxiv.org/pdf/cond-mat/0512155.pdf} (\bibinfo {year} {2005})\BibitemShut {NoStop}%
\bibitem [{\citenamefont {Harris}\ \emph {et~al.}(2012)\citenamefont {Harris}, \citenamefont {Peter}, \citenamefont {Bellis}, \citenamefont {Baum}, \citenamefont {Kabla},\ and\ \citenamefont {Charras}}]{charras_cell_1}%
  \BibitemOpen
  \bibfield  {author} {\bibinfo {author} {\bibfnamefont {A.~R.}\ \bibnamefont {Harris}}, \bibinfo {author} {\bibfnamefont {L.}~\bibnamefont {Peter}}, \bibinfo {author} {\bibfnamefont {J.}~\bibnamefont {Bellis}}, \bibinfo {author} {\bibfnamefont {B.}~\bibnamefont {Baum}}, \bibinfo {author} {\bibfnamefont {A.~J.}\ \bibnamefont {Kabla}},\ and\ \bibinfo {author} {\bibfnamefont {G.~T.}\ \bibnamefont {Charras}},\ }\href {https://doi.org/https://doi.org/10.1073/pnas.1213301109} {\bibfield  {journal} {\bibinfo  {journal} {Proceedings of the National Academy of Sciences}\ }\textbf {\bibinfo {volume} {41}},\ \bibinfo {pages} {16449} (\bibinfo {year} {2012})}\BibitemShut {NoStop}%
\bibitem [{\citenamefont {Khalilgharibi}\ \emph {et~al.}(2019)\citenamefont {Khalilgharibi}, \citenamefont {Fouchard}, \citenamefont {Asadipour}, \citenamefont {Barrientos}, \citenamefont {Duda}, \citenamefont {Bonfanti}, \citenamefont {Yonis}, \citenamefont {Harris}, \citenamefont {Mosaffa}, \citenamefont {Fujita}, \citenamefont {Kabla}, \citenamefont {Mao}, \citenamefont {Baum}, \citenamefont {Mu{\~{n}}oz}, \citenamefont {Miodownik},\ and\ \citenamefont {Charras}}]{charras_cell_2}%
  \BibitemOpen
  \bibfield  {author} {\bibinfo {author} {\bibfnamefont {N.}~\bibnamefont {Khalilgharibi}}, \bibinfo {author} {\bibfnamefont {J.}~\bibnamefont {Fouchard}}, \bibinfo {author} {\bibfnamefont {N.}~\bibnamefont {Asadipour}}, \bibinfo {author} {\bibfnamefont {R.}~\bibnamefont {Barrientos}}, \bibinfo {author} {\bibfnamefont {M.}~\bibnamefont {Duda}}, \bibinfo {author} {\bibfnamefont {A.}~\bibnamefont {Bonfanti}}, \bibinfo {author} {\bibfnamefont {A.}~\bibnamefont {Yonis}}, \bibinfo {author} {\bibfnamefont {A.}~\bibnamefont {Harris}}, \bibinfo {author} {\bibfnamefont {P.}~\bibnamefont {Mosaffa}}, \bibinfo {author} {\bibfnamefont {Y.}~\bibnamefont {Fujita}}, \bibinfo {author} {\bibfnamefont {A.}~\bibnamefont {Kabla}}, \bibinfo {author} {\bibfnamefont {Y.}~\bibnamefont {Mao}}, \bibinfo {author} {\bibfnamefont {B.}~\bibnamefont {Baum}}, \bibinfo {author} {\bibfnamefont {J.~J.}\ \bibnamefont {Mu{\~{n}}oz}}, \bibinfo {author} {\bibfnamefont {M.}~\bibnamefont {Miodownik}},\ and\ \bibinfo {author} {\bibfnamefont
  {G.}~\bibnamefont {Charras}},\ }\href {https://doi.org/10.1038/s41567-019-0516-6} {\bibfield  {journal} {\bibinfo  {journal} {Nature Physics}\ }\textbf {\bibinfo {volume} {15}},\ \bibinfo {pages} {839} (\bibinfo {year} {2019})}\BibitemShut {NoStop}%
\bibitem [{\citenamefont {Goren}\ \emph {et~al.}(2023)\citenamefont {Goren}, \citenamefont {Levin}, \citenamefont {Brand}, \citenamefont {Lesman},\ and\ \citenamefont {Sorkin}}]{goren2023probing}%
  \BibitemOpen
  \bibfield  {author} {\bibinfo {author} {\bibfnamefont {S.}~\bibnamefont {Goren}}, \bibinfo {author} {\bibfnamefont {M.}~\bibnamefont {Levin}}, \bibinfo {author} {\bibfnamefont {G.}~\bibnamefont {Brand}}, \bibinfo {author} {\bibfnamefont {A.}~\bibnamefont {Lesman}},\ and\ \bibinfo {author} {\bibfnamefont {R.}~\bibnamefont {Sorkin}},\ }\href {https://doi.org/https://doi.org/10.1002/smll.202202573} {\bibfield  {journal} {\bibinfo  {journal} {Small}\ }\textbf {\bibinfo {volume} {19}},\ \bibinfo {pages} {2202573} (\bibinfo {year} {2023})}\BibitemShut {NoStop}%
\bibitem [{\citenamefont {Bramwell}\ and\ \citenamefont {Gingras}(2001)}]{spin_liquid}%
  \BibitemOpen
  \bibfield  {author} {\bibinfo {author} {\bibfnamefont {S.~T.}\ \bibnamefont {Bramwell}}\ and\ \bibinfo {author} {\bibfnamefont {M.~J.~P.}\ \bibnamefont {Gingras}},\ }\href {https://doi.org/10.1126/science.1064761} {\bibfield  {journal} {\bibinfo  {journal} {Science}\ }\textbf {\bibinfo {volume} {294}},\ \bibinfo {pages} {1495} (\bibinfo {year} {2001})}\BibitemShut {NoStop}%
\bibitem [{\citenamefont {Livne}\ \emph {et~al.}(2023)\citenamefont {Livne}, \citenamefont {Schiller},\ and\ \citenamefont {Moshe}}]{Livne:2023aa}%
  \BibitemOpen
  \bibfield  {author} {\bibinfo {author} {\bibfnamefont {N.~S.}\ \bibnamefont {Livne}}, \bibinfo {author} {\bibfnamefont {A.}~\bibnamefont {Schiller}},\ and\ \bibinfo {author} {\bibfnamefont {M.}~\bibnamefont {Moshe}},\ }\href {https://doi.org/10.1103/PhysRevE.107.055004} {\bibfield  {journal} {\bibinfo  {journal} {Physical Review E}\ }\textbf {\bibinfo {volume} {107}},\ \bibinfo {pages} {055004} (\bibinfo {year} {2023})}\BibitemShut {NoStop}%
\bibitem [{\citenamefont {Ortiz-Tav\'arez}\ \emph {et~al.}(2025)\citenamefont {Ortiz-Tav\'arez}, \citenamefont {Yang}, \citenamefont {Kotov},\ and\ \citenamefont {Mao}}]{PhysRevLett.134.147401}%
  \BibitemOpen
  \bibfield  {author} {\bibinfo {author} {\bibfnamefont {J.~M.}\ \bibnamefont {Ortiz-Tav\'arez}}, \bibinfo {author} {\bibfnamefont {Z.}~\bibnamefont {Yang}}, \bibinfo {author} {\bibfnamefont {N.}~\bibnamefont {Kotov}},\ and\ \bibinfo {author} {\bibfnamefont {X.}~\bibnamefont {Mao}},\ }\href {https://doi.org/10.1103/PhysRevLett.134.147401} {\bibfield  {journal} {\bibinfo  {journal} {Physical Review Letters}\ }\textbf {\bibinfo {volume} {134}},\ \bibinfo {pages} {147401} (\bibinfo {year} {2025})}\BibitemShut {NoStop}%
\bibitem [{\citenamefont {Zhang}\ \emph {et~al.}(2019)\citenamefont {Zhang}, \citenamefont {Zhang}, \citenamefont {Bouzid}, \citenamefont {Rocklin}, \citenamefont {Del~Gado},\ and\ \citenamefont {Mao}}]{PhysRevLett.123.058001}%
  \BibitemOpen
  \bibfield  {author} {\bibinfo {author} {\bibfnamefont {S.}~\bibnamefont {Zhang}}, \bibinfo {author} {\bibfnamefont {L.}~\bibnamefont {Zhang}}, \bibinfo {author} {\bibfnamefont {M.}~\bibnamefont {Bouzid}}, \bibinfo {author} {\bibfnamefont {D.~Z.}\ \bibnamefont {Rocklin}}, \bibinfo {author} {\bibfnamefont {E.}~\bibnamefont {Del~Gado}},\ and\ \bibinfo {author} {\bibfnamefont {X.}~\bibnamefont {Mao}},\ }\href {https://doi.org/10.1103/PhysRevLett.123.058001} {\bibfield  {journal} {\bibinfo  {journal} {Physical Review Letter}\ }\textbf {\bibinfo {volume} {123}},\ \bibinfo {pages} {058001} (\bibinfo {year} {2019})}\BibitemShut {NoStop}%
\bibitem [{\citenamefont {Benoist}\ \emph {et~al.}(2023)\citenamefont {Benoist}, \citenamefont {Saggiorato},\ and\ \citenamefont {Lenz}}]{benoist2023generic}%
  \BibitemOpen
  \bibfield  {author} {\bibinfo {author} {\bibfnamefont {F.}~\bibnamefont {Benoist}}, \bibinfo {author} {\bibfnamefont {G.}~\bibnamefont {Saggiorato}},\ and\ \bibinfo {author} {\bibfnamefont {M.}~\bibnamefont {Lenz}},\ }\href {https://doi.org/10.1039/d2sm01606k} {\bibfield  {journal} {\bibinfo  {journal} {Soft Matter}\ }\textbf {\bibinfo {volume} {19}},\ \bibinfo {pages} {2970} (\bibinfo {year} {2023})}\BibitemShut {NoStop}%
\bibitem [{\citenamefont {Obolensky}\ \emph {et~al.}(2009)\citenamefont {Obolensky}, \citenamefont {Doerr}, \citenamefont {Ray},\ and\ \citenamefont {Yu}}]{obolensky2009rigorous}%
  \BibitemOpen
  \bibfield  {author} {\bibinfo {author} {\bibfnamefont {O.}~\bibnamefont {Obolensky}}, \bibinfo {author} {\bibfnamefont {T.}~\bibnamefont {Doerr}}, \bibinfo {author} {\bibfnamefont {R.}~\bibnamefont {Ray}},\ and\ \bibinfo {author} {\bibfnamefont {Y.-K.}\ \bibnamefont {Yu}},\ }\href {https://doi.org/10.1103/PhysRevE.79.041907} {\bibfield  {journal} {\bibinfo  {journal} {Physical Review E}\ }\textbf {\bibinfo {volume} {79}},\ \bibinfo {pages} {041907} (\bibinfo {year} {2009})}\BibitemShut {NoStop}%
\end{thebibliography}%



\clearpage\newpage

\end{document}